\documentclass[12pt,a4paper]{article}
\pdfoutput=1
\usepackage{jheppub}
\usepackage{epstopdf}
\usepackage{graphicx}
\usepackage{epsfig}
\usepackage{dcolumn}  
\usepackage{bm}    
\usepackage{amssymb} 
\usepackage{amsmath,bm}
\usepackage{amsfonts}    
\usepackage{slashed}  
\usepackage[mathscr]{euscript}
\usepackage{epsfig}
\usepackage[position=t,singlelinecheck=off]{subfig}
\usepackage{caption}
\title{Probing crunching AdS cosmologies }

\author{S. Prem Kumar} 
\author{and Vladislav Vaganov} 
\affiliation{Department of Physics, Swansea University,\\Singleton Park, Swansea SA2 8PP, UK.}
\emailAdd{s.p.kumar, pyvv@swansea.ac.uk}
\abstract{Holographic gravity duals of deformations of CFTs formulated on de Sitter spacetime contain FRW geometries behind a horizon, with cosmological big crunch singularities. Using a specific analytically tractable solution within a particular single scalar truncation of ${\cal N}=8$ supergravity on AdS$_4$,
we first probe such crunching cosmologies with spacelike radial geodesics that compute spatially antipodal correlators of large dimension boundary operators. At late times, the geodesics lie on the FRW slice of maximal expansion behind the horizon. The late time two-point functions factorise, and when transformed to the Einstein static universe, they exhibit a temporal non-analyticity  determined by the maximal value of the scale factor  $\tilde a_{\rm max}$.  Radial geodesics connecting antipodal points necessarily have de Sitter energy ${\cal E}\lesssim \tilde a_{\rm max}$, while geodesics with ${\cal E}> \tilde a_{\rm max}$ terminate at the crunch, the two categories of geodesics being separated by the maximal expansion slice.
The spacelike crunch singularity is curved ``outward'' in the Penrose diagram for the deformed AdS backgrounds, and thus geodesic limits of  the antipodal correlators do not directly probe the crunch. Beyond the geodesic limit, we point out that the scalar wave equation, analytically continued into the FRW patch, has a potential which is singular at the crunch along with complex WKB turning points in the vicinity of the FRW crunch. We then argue that the frequency space Green's function has a branch point determined by $\tilde a_{\rm max}$ which corresponds to the lowest quasinormal frequency. 
}

\begin{document}
 \maketitle 

\def\be{\begin{equation}}
\def\ee{\end{equation}}
\def\bea{\begin{eqnarray}} 
\def\eea{\end{eqnarray}}
\def\nn{\nonumber}
\def\pd{\partial}
\def\Re{R\'{e}nyi }
\def\l1{{\text{1-loop}}}
\def\uy{u_y}
\def\ur{u_R}
\def\o{\mathcal{O}}
\def\Cl{{{cl}}}
\def\bz{{\bar{z}}}
\def\by{{\bar{y}}}
\def\bX{\bar{X}}
\def\im{{\text{Im}}}
\def\re{{\text{Re}}}
\def\cn{{\text{cn}}}
\def\sn{{\text{sn}}}
\def\dn{{\text{dn}}}
\def\K{\mathbf{K}}
\def\n1{\Bigg|_{n=1}}
\def\fin{{\text{finite}}}
\def\R{{\mathscr{R}}}
\def\one{{(1)}}
\def\zero{{(0)}}
\def\n{{(n)}}
\def\tr{\text{Tr}}
\def\T{\mathcal{T}}
\def\TT{\tilde{\mathcal{T}}}
\def\O{\mathcal{O}}
\def\cN{\mathcal{N}}
\def\P{\Phi}
\def\cosech{{\text{cosech}}}
\def\csch{{\text{cosech}}}
\def\W{{\tilde{W}}}
\def\T{{\tilde{T}}}
\def\sgn{{\rm sgn}}
\def\by{\bar{y}}
\def\arcsinh{\sinh^{-1}}
\newcommand*\xbar[1]{%
  \hbox{%
    \vbox{%
      \hrule height 0.5pt % The actual bar
      \kern0.5ex%         % Distance between bar and symbol
      \hbox{%
        \kern-0.1em%      % Shortening on the left side
        \ensuremath{#1}%
        \kern-0.1em%      % Shortening on the right side
      }%
    }%
  }%
}

\section{Introduction and Summary}
Understanding the physics in the vicinity of spacelike singularities within a well-defined microscopic framework for gravity is one of the outstanding challenges in theoretical physics. The AdS/CFT correspondence and holography \cite{maldacena, witten} provide some of the most promising routes for exploring this and related questions. Several important advances in this direction have been made,  within the general framework of strings and holography, e.g. \cite{Craps:2006xq, Das:2006dz, Awad:2008jf}, and through the embedding of cosmological crunching geometries in Anti-de Sitter (AdS) spacetime \cite{HH1, HH2, Turok:2007ry, Craps:2007ch}, and more recently in \cite{EHH1, EHH2}. The main goal in a holographic setting of this type is to find out how a given bulk (cosmological) singularity makes itself known within the dual quantum field theory. The answer to this question (at strong coupling and/or large $N$) could then be used to eventually understand  a potential resolution of the bulk singularity  via the specific  holographic dictionary (as stringy and/or quantum effects). 

While bulk singularities may or may not be cloaked  behind a horizon, one expects the information on the singularity to be encoded in some way within  correlation functions of the boundary field theory. For the black hole singularity in AdS, it was shown in \cite{shenkeretal} that two-point thermal correlators of large dimension boundary operators (inserted on the two boundaries of AdS-Schwarzschild), computed by  spacelike bulk geodesics, subtly encode information on the black hole singularity via a temporal non-analyticity. This was expected primarily due to the fact that the geodesics in question penetrate the horizon and can get arbitrarily close to the black hole singularity. The implication of the behaviour of such probe geodesics for frequency space (thermal) Wightman functions was worked out precisely in \cite{fl1}
applying a WKB approximation to wave equations in the bulk. The general idea of probing bulk singularities with correlators in the geodesic limit has proved to be extremely attractive and has been applied recently to deduce a direct signature of singularities in anisotropic Kasner-AdS solutions within the holographic setting \cite{EHH1, EHH2, Banerjee:2015fua}.

Cosmological singularities have also found a natural place within the AdS/CFT setting following the works of \cite{HH1, HH2}. Such crunching AdS cosmologies arise naturally as FRW geometries (with spatially hyperbolic sections) behind bulk horizons within gravity duals of deformed CFTs placed in de Sitter (dS) spacetime \cite{maldacena1, Harlow:2010my}. Following an appropriate boundary conformal transformation, the same setup can be viewed as the gravity dual of a deformed CFT on the boundary (${\mathbb R}\times S^{d-1}$ or Einstein static universe) of global  AdS spacetime. In this latter picture, the bulk singularity hits the boundary at finite global time, at which point the field theory evolution has been argued to be singular 
\cite{HH2, Turok:2007ry, Craps:2007ch, Barbon:2011ta}. On general grounds, the  field theoretic ``singularity'' in the second picture (Einstein static universe) has been related to time dependent couplings ``driving'' various  condensates to diverge at a finite value of the global time \cite{Barbon:2011ta, Barbon:2013nta}. In contrast, the field theoretic evolution in the de Sitter frame picture can be perfectly smooth and well defined for appropriately chosen boundary conditions and sufficiently small (relevant) deformations, as was remarked in \cite{maldacena1}.

Given that thermal field theory correlators potentially encode (upon appropriate continuation) certain aspects of the black hole geometry behind the horizon \cite{shenkeretal, fl1, fl2}, it is natural to ask in what sense correlation functions of de Sitter space field theories probe the FRW crunching patches behind the horizon of the dual gravity theories\footnote{The question is inevitably posed in the strong coupling limit of the field theory and  gauge/gravity duality  used to  identify and compute the relevant boundary observables. This paves the way, at least in principle, for eventually investigating  ideas such as singularity resolution from the field theory side of the gauge/gravity duality.}.  The aim of this paper is to initiate a study of correlators within strongly interacting deformed CFTs on de Sitter spacetime with a view towards identifying certain aspects of observables that have knowledge of the FRW geometry behind the bulk horizon. It is known that  extremal surfaces of the kind that probe the FRW patch will generically be bounded by some extremal slice away from the crunch singularity 
\cite{hubeny, maldacena2, Fischler:2013fba, engelhardt}. This makes the problem of examining the appropriate holographic correlators more challenging, and therefore worthy of study. In particular, one would need to understand the issues involved in going beyond the geodesic limit employing WKB-like approximations (as in \cite{fl1, fl2}) and connecting these to well-defined field theoretic observables in de Sitter spacetime. Importantly, in order to be in a position to identify the kinds of subtle features that were uncovered for the AdS-Schwarzschild black hole, we also require an analytically tractable gravity background exhibiting a genuine cosmological crunch in the FRW patch.

We first focus our attention on the two-point correlator in the boundary field theory between spatially antipodal points (the North and South poles) of global de Sitter spacetime. The two points remain causally disconnected for all times (see e.g. \cite{Spradlin:2001pw}), and in this sense the antipodal correlator is similar to the AdS black hole geodesic correlator across the two boundaries. A key difference between the two systems is that the dS-sliced, asymptotically AdS backgrounds possess only one boundary, and their Penrose diagrams when drawn with a ``straightened'' conformal boundary, render the spacelike  crunch singularity curved ``outwards'' (in contrast to AdS-Schwarzschild \cite{shenkeretal}).  Spacelike geodesics which are radially directed in dS-sliced backgrounds also possess a conserved ``energy'' ${\cal E}$ which can eventually be identified with imaginary  de Sitter frequency  divided by the conformal dimension $\Delta$ of the operator in  the boundary field theory. 

We study in explicit detail the properties of the antipodal correlator in the geodesic approximation and at late times, and make contact with the massive scalar wave equation in the WKB limit, continued into the FRW patch. Whilst our analysis is general, for concreteness we also focus attention on a specific analytically tractable deformation of AdS$_4$ obtained within a particular single scalar truncation of ${\cal N}=8$ supergravity \cite{yiannis0, yiannis2, yiannis1}. The scalar in question has mass squared  $-2$ in units of the AdS radius, exactly as the truncation considered in the original works on crunching AdS cosmologies \cite{HH1, HH2}, therefore permitting two inequivalent quantisations. The scalar potentials in the two different truncations are, however, different. The exact solution presented in \cite{Papadimitriou:2004ap, yiannis1} yields a smooth deformation of Euclidean AdS$_4$ which we continue to Lorentzian signature in order to obtain the dS-sliced deformed background and an FRW patch with hyperbolic ($H^3$) slices behind a horizon. 

Our main results and observations can be summarised as follows:
\begin{itemize}
\item{As expected on general grounds, antipodal geodesics probing the FRW patch remain bounded by the slice of maximal expansion with FRW scale factor $\tilde a_{\rm max}$. In particular, in the late time limit, geodesics lie almost entirely on the slice of maximal expansion and the corresponding correlator behaves as,
\bea
&&\langle{\cal O}_{\Delta}(\tau_1,0)\,{\cal O}_{\Delta}(\tau_2,\,\pi) \rangle_{\rm ESU}\,\sim\, \left(\tfrac{\pi}{2}-\tau_1\right)^{\Delta(\tilde a_{\rm max}-1)}\,\left(\tfrac{\pi}{2}-\tau_2\right)^{\Delta(\tilde a_{\rm max}-1)}\,\\\nonumber\\\nonumber
&&\langle{\cal O}_{\Delta}(t_1,0)\,{\cal O}_{\Delta}(t_2,\,\pi) \rangle_{\rm dS}\,\sim\, e^{-\tilde a_{\rm max}(t_1+t_2)\Delta}\,,
\eea
in the Einstein static and de Sitter frames, respectively. Undeformed AdS has $\tilde a_{\rm max}=1$ and therefore the limit $\tau_{1,2}\to \tfrac{\pi}{2}$ is smooth. In the presence of a deformation, $\tilde a_{\rm max} < 1$, and  the correlator on ${\mathbb R}\times S^{d-1}$ exhibits non-analytic behaviour  when the bulk singularity hits the boundary at finite time. In contrast the dS-space correlators are always smooth. A second inference one may draw from the above behaviour is that at late times, correlators appear to factorise, suggesting the presence of a dominant, disconnected contribution i.e.  growing condensates. This is in line with the general arguments of \cite{Barbon:2011ta} associating the crunch singularity at the boundary to a ``CFT fall'' i.e. homogeneous condensates diverging at a finite time in the Einstein static frame.}
\item{The antipodal, radial geodesics have an associated conserved de Sitter energy ${\cal E}$ which is bounded from above by $\tilde a_{\rm max}$. In particular, spacelike geodesics with ${\cal E} > \tilde a_{\rm max}$ penetrate the bulk horizon and terminate at the crunch. Performing a de Sitter mode expansion for a massive probe scalar in the bulk, the wave equation in the WKB limit coincides with the geodesic equation, upon making the identification ${\cal E}=-i\nu/\Delta$, where $\nu$ denotes the frequency of appropriately defined de Sitter modes.  The absence of geodesics with two boundary endpoints for ${\cal E}> \tilde a_{\rm max}$ suggests the possibility of a corresponding feature in appropriately defined frequency space correlators.
}
\item{ The wave equation in the WKB limit exhibits nontrivial turning points upon analytic continuation into the FRW patch. One of these is the continuation of the unique WKB turning point outside the horizon. The merger of this turning point with a complex turning point gives rise to branch points in WKB correlators. We make this statement precise for the retarded, frequency space correlator on the boundary. In particular, the branch point for purely imaginary frequency is  associated to the position of the lowest quasinormal pole \cite{fl2}, and the corresponding branch cut has a natural interpretation  in terms of discrete quasinormal modes forming a continuum in the scaling limit required for implementing WKB. The singularity in the Schr\"odinger potential,  however, suggests further interesting physics.}

\item{The scalar wave equation, when expressed in Schr\"odinger form and continued into the FRW patch, exhibits a singular potential precisely at the crunch. In particular, it universally diverges as the inverse square of the (tortoise) coordinate distance to the crunch. The coefficient and the sign of the divergent contribution are model dependent. For the AdS$_4$ deformation studied in this paper, the potential diverges to negative infinity at the location of the crunch. This leading singularity in the potential is  formally subdominant in  the WKB limit indicating that a proper inclusion of its effects will become necessary to obtain the correct description of boundary correlators for complex frequencies.
}
\end{itemize}

The layout of the paper is as follows: in section 2 we collect together basic aspects of the crunching AdS backgrounds obtained by continuation from smooth Euclidean solutions. Section 3 deals with the general features of antipodal geodesics and their explicit solutions in AdS spacetime, both in global coordinates and in the dS-sliced picture. In Section 4, we provide a detailed description of the analytically tractable crunching model in deformed AdS$_4$. We plot the antipodal geodesics numerically and examine their late time behaviour. Section 5 is aimed at making contact with the WKB limit of the wave equation and properties of the Schr\"odinger potential near the crunch. Finally, in the appendix, we collect together useful transformations, and properties of de Sitter mode expansions; appendix C  presents a detailed  holographic derivation of the position-space antipodal correlator in a WKB-like limit in Euclidean AdS$_{d+1}$.

\section{Euclidean AdS instantons and crunches}
We begin by recalling certain basic elements of de Sitter-sliced, asymptotically AdS space times. The main point of this section is to lay out 
coordinates and conventions appropriate for analytic continuation into the FRW patch, and to explain the shape of the spacelike crunch singularity in the corresponding Penrose diagram.

\label{sec:intro}
Asymptotically AdS spacetimes containing FRW crunches can be obtained by  appropriate analytic continuation of asymptotically Euclidean-AdS$_{d+1}$ (EAdS$_{d+1}$) geometries with the $d$-dimensional sphere as conformal boundary (see e.g. \cite{maldacena1,maldacena2}). The metric for an asymptotically EAdS$_{d+1}$ geometry, with the topology of a ball, can be taken to be of the form
\be
ds^2_{\rm E}\,=\, d\xi^2\,+\,a(\xi)^2\,{d\Omega_{d}^2}\,,\qquad\qquad 0\leq \xi<\infty\,,
\ee
along with the requirement of AdS asymptotics and smoothness at the origin
\be
a(\xi)\left.\right|_{\xi\to\infty}\,\sim\,e^{\xi}\,,\qquad\qquad a(\xi)\left.\right|_{\xi\to 0}\,\simeq\,\xi\,.
\ee
Anti-de Sitter spacetime is obtained when 
\be
a(\xi)\,=\,a_{\rm AdS}(\xi)\,=\sinh\xi\,,
\ee
setting the AdS radius to unity.
Such Euclidean AdS ``instantons'' have a natural place in the study of vacuum decay in a theory of gravity (see e.g. \cite{cdl1, maldacena1}). Within the context of AdS/CFT duality, classical (super)gravity in the asymptotically EAdS$_{d+1}$ background then computes the observables of a CFT on the conformal boundary which is the $d$-dimensional sphere S$^{d}$. Upon appropriate analytic continuation to Lorentzian signature this setup relates the CFT on a fixed de Sitter spacetime dS$_d$ to  gravity in AdS$_{d+1}$: 
\bea
&&d\Omega_d^2\,=\,d\theta^2 \,+\,\sin^2\theta\,d\Omega_{d-1}^2\,, 
\\\nonumber
\\
&&\theta \to it + \frac{\pi}{2}\,,
\qquad d\Omega_d^2\to -\,dt^2\,+\,\cosh^2 t\,d\Omega_{d-1}^2\,.
\label{cont}
\eea
With this continuation, the origin $\xi=0$ becomes the lightcone or horizon of the Lorentzian solution which splits the asymptotically AdS$_{d+1}$ spacetime into two coordinate patches:

\paragraph{Exterior region (I):} The exterior ``bubble" or ``Euclidean" region, outside the lightcone.   This is the straightforward continuation \eqref{cont} of the Euclidean AdS instanton, and corresponds to a patch of the asymptotically AdS$_{d+1}$ spacetime with global de Sitter (dS$_d$) slices with $SO(d,1)$ isometry: 
\begin{equation}
ds^2\, =\, d\xi^2\, +\, a^2(\xi)\left(-dt^2 \,+\,\cosh^2{t}\,d\Omega_{d-1}^2\right)\,.
\end{equation}
The coordinate ranges are $0\leq\xi<\infty$ and $-\infty \leq t < \infty$.
This region has $a(\xi)^2\geq 0$, vanishing as $a^2\simeq \xi^2$ at $\xi=0$.

\paragraph{ FRW region (II):}  The interior ``FRW'' or ``cosmological'' region inside the lightcone at $\xi=0$.   To get to this patch we analytically continue the 
coordinates and scale factor in region I as:
\bea
 \xi\to i\sigma\,,\qquad t\to \chi\,-\,\frac{i\pi}{2}\,,\qquad a\to i\, \tilde{a}\,.
 \label{toFRW}
 \eea
This yields an FRW universe with hyperbolic spatial slices ${\mathbb H}_d$:
\begin{equation}
ds^2\, =\, -d\sigma^2\, +\, \tilde{a}^2(\sigma)(d\chi^2\, +\,\sinh^2{\chi}\,d\Omega_{d-1}^2)\,.\label{region1}
\end{equation}
The $SO(d,1)$ symmetry now acts on the ${\mathbb H}_d$ slices.
The radial coordinate $\chi$ along the non-compact spatial slices has infinite range, while the temporal coordinate $\sigma$ is finite and bounded by a crunch singularity at $\sigma = \sigma_c$ where $\tilde a$ has a second zero:
\be
0\leq \chi <\infty\,,\qquad 0\leq \sigma < \sigma_c\,.
\ee
For the undeformed AdS$_{d+1}$ spacetime $\tilde a (\sigma)\,=\,\sin\sigma$, and although the FRW scale factor vanishes at $\sigma = \sigma_c =\pi$ this is only a coordinate singularity since AdS is nonsingular with constant curvature $R = - d(d+1)/L^2_{\rm AdS}$ (we implicitly set the AdS radius $L_{\rm AdS} =1$). However, the situation changes in the presence of any deformation, relevant or otherwise. The interior FRW region arises naturally in the study of vacuum decay {a la} Coleman and de Luccia \cite{cdl1}.
As originally argued in \cite{cdl1, cdl2}, any deformation away from AdS will result in $\tilde a(\sigma)$ first increasing from $\tilde a(0)=0$, reaching a maximum and subsequently decreasing to zero at a big crunch singularity at $\sigma = \sigma_c$. The inevitability of such a singularity follows from the existence of a closed trapped surface in the crunching regime (as argued by Abbott and Coleman \cite{cdl2}). 

\subsection{Conformal Structure}

The deformations of AdS$_{d+1}$ that we are interested in have the property that they can always be expressed as a conformal factor times the undeformed EAdS metric. This is clear in Euclidean signature following a coordinate change,
\bea
&& ds^2\,=\,d\xi^2\,+\,a(\xi)^2\,d\Omega_d^2\,\,=\,\,\Lambda(\hat \xi)^2\,\left(d\hat\xi^2\,+\,\sinh^2\hat\xi\,d\Omega_d^2\,\right)\,,
\eea
where $\hat\xi$ and the conformal factor $\Lambda$ are determined by solving the differential equations:
\be
\hat \xi^{\,\prime}(\xi)\,=\,
\frac{\sinh\hat \xi}{a(\xi)}\,,\qquad\qquad \Lambda\,=\,\frac{1}{\hat\xi^{\,\prime}(\xi)}\,\,.
\ee
The radial coordinate $\hat \xi$ of the EAdS metric lies in the range $0\leq \hat\xi <\infty$ with $\hat \xi =0$ being the origin. Now the full metric can be continued to Lorentzian signature as in the preceding discussion leading to an exterior region I:
\be
ds^2\,=\,\Lambda(\hat\xi)^2\left[d\hat\xi^2\,+\,\sinh^2\hat \xi\left(-\,dt^2\,+\,\cosh^2 t\,d\Omega_{d-1}^2\right)\right]\,,
\ee
and the interior FRW patch, or region II:
\bea
&&ds^2\,=\,\tilde\Lambda(\hat\sigma)^2\left[-d\hat\sigma^2\,+\,\sin^2\hat\sigma\left(d\chi^2\,+\,\sinh^2 \chi\,d\Omega_{d-1}^2\right)\right]\,,\\\nonumber\\\nonumber
&&\tilde\Lambda(\hat\sigma)\,\equiv\,\Lambda(i\hat \sigma)\,.
\eea
Therefore, the conformal (Penrose) diagram for these spacetimes is the same as AdS, except that the range of the FRW time in region II is limited by the crunch singularity at some $\hat\sigma\,=\,\hat \sigma_c$ where the analytically continued conformal factor vanishes:
\be
0\leq \hat \sigma \leq \hat\sigma_c\,,\qquad \ \tilde\Lambda(\hat\sigma_c)\,=\,0\,.
\ee
\begin{figure}
\centering
\includegraphics[width= 2in]{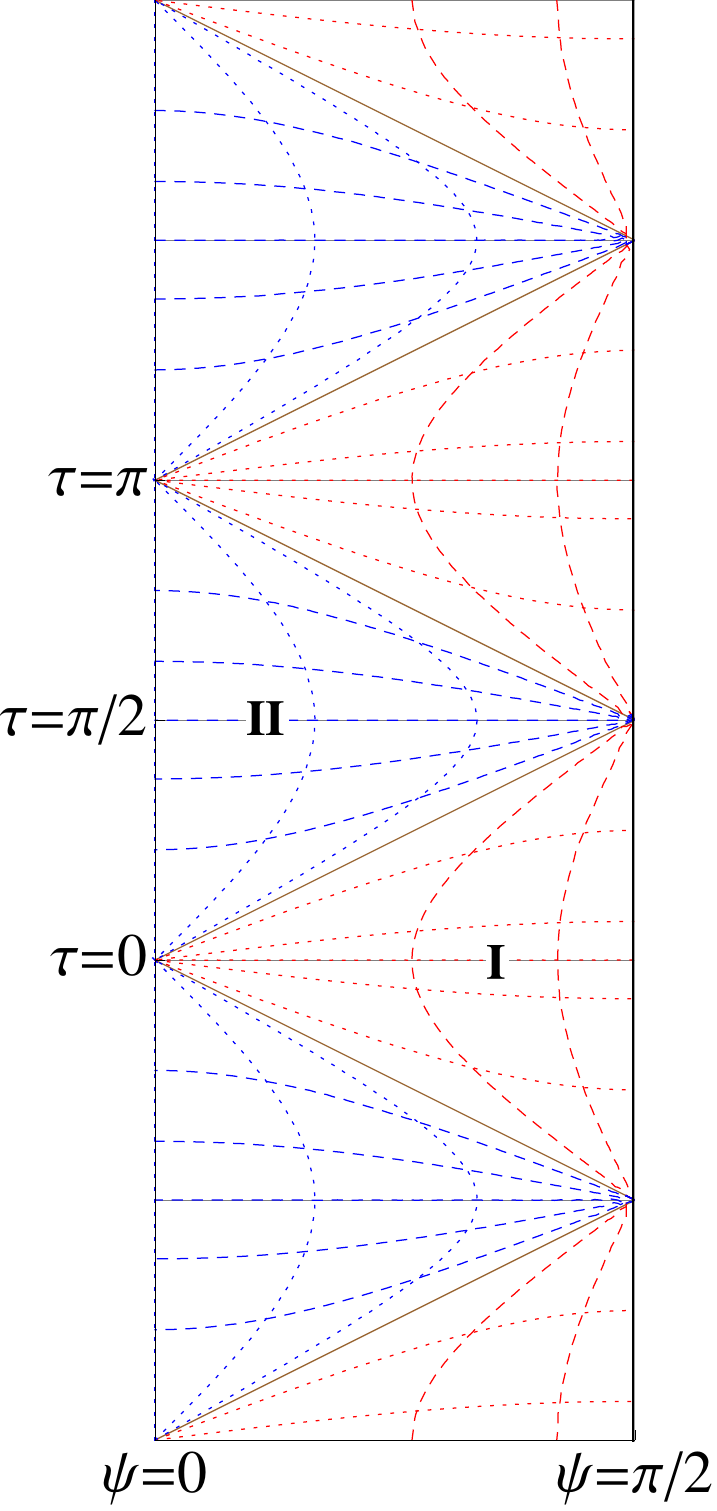}\hspace{1in}
\includegraphics[width=2.2in]{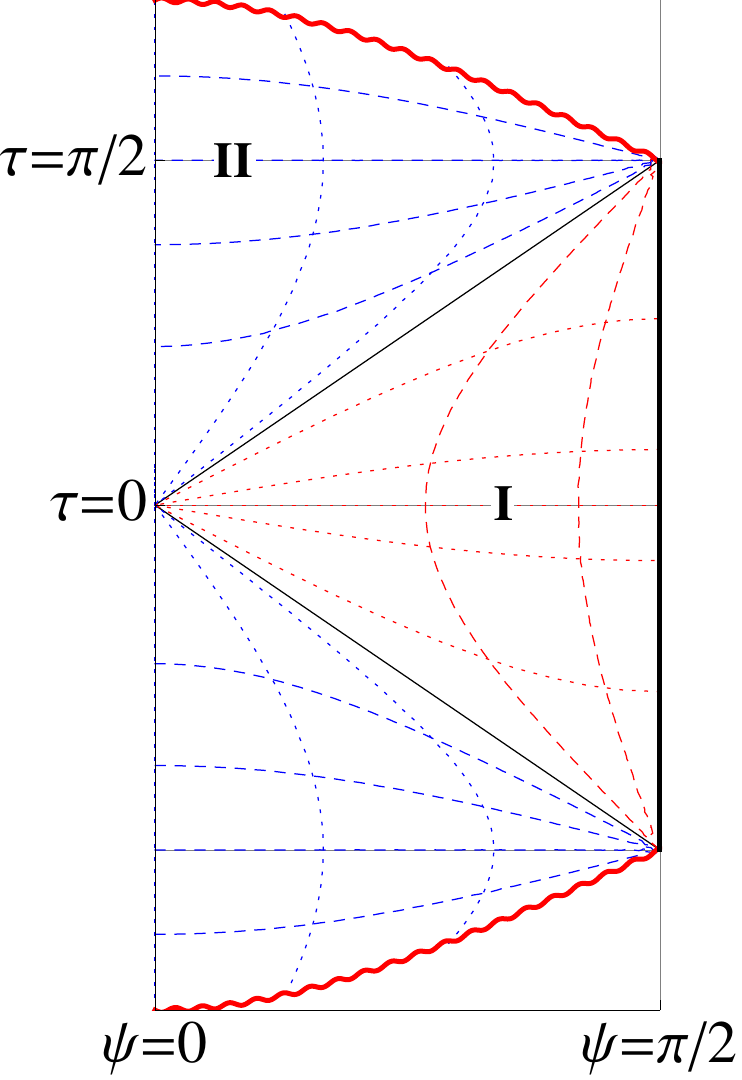}\vspace{0.2in}
\caption{\small{{\bf Left:} Penrose diagram of AdS$_{d+1}$. 
{\bf Right:} Penrose diagram of deformed AdS$_{d+1}$ with the big crunch singularity represented by the jagged red line. 
In both diagrams the conformal boundary is at $\psi=\pi/2$, each point representing a sphere S$^{d-1}$.   Horizontal lines are contours of constant global time $\tau$ while vertical lines are contours of constant global radial coordinate $r=\tan{\psi}$. In the exterior region I, dashed red lines are contours of constant $\hat{\xi}$, whilst dotted red lines are contours of constant $t$.   In the FRW region II, dashed blue lines are contours of constant $\hat{\sigma}$, dotted blue lines are contours of constant $\chi$.}}
\label{fig:pda}
\end{figure}
The Penrose diagram of AdS$_{d+1}$ is shown in figure \ref{fig:pda}. This is obtained after transformation to global coordinates (see appendix \ref{app:dstoglobal} for details), so that the conformal boundary is naturally the cylinder ${\mathbb R}\times S^{d-1}$. The radial coordinate $\psi$ ranges from $0$ to $\pi/2$, and $\tau$ is the global time.
Region I is marked with contours of constant $\hat{\xi}$ and $t$ in red, while region II is covered by the FRW coordinates with contours of constant $\hat{\sigma}$ and $\chi$ in blue.    The diagram for undeformed AdS has a periodicity in global time $\tau$, given by $\Delta\tau=2\pi$, 
resulting in an infinite number of such patches stacked on top of each other.   

In contrast, the Penrose diagram of deformed AdS$_{d+1}$ in figure \ref{fig:pda} shows a jagged line indicating the spacelike  big crunch singularity at which spacetime ends, and  cannot be extended any further. Hence the diagram has only a single copy of the exterior and the interior FRW patches for $\tau \geq 0$ and is no longer periodic.

Although the Penrose diagram bears similarities to the Schwarzschild-AdS black hole\footnote{Particularly if we include its mirror image about the vertical axis corresponding to negative azimuthal angles.}, it is different in two crucial ways:
\begin{itemize}
\item
{There is only one asymptotic region unlike Schwarzschild-AdS which has two. This observation is an immediate consequence of the fact that the deformed spacetime is conformal to AdS.}
\item
{If the boundary is drawn straight, as shown, the singularity is curved \emph{outwards} with respect to the horizontal $\tau=\pi/2$.   For Schwarzschild-AdS$_{d+1}$, it was shown in   \cite{shenkeretal} that the singularity is curved \emph{inwards}\footnote{Provided $d>2$; in the $d=2$ case, the BTZ black hole, the diagram is square.}. We will justify this statement below by examining the geometry using a Kruskal-like extension.}
\end{itemize}

\subsection{Kruskal-like extension}
The Lorentzian geometries with horizons permit natural Kruskal-like coordinates that can describe both regions I and II. Suppressing the angular direction, the relevant pieces of the metric  in the $t-\xi$ section in Lorentzian signature are 
\begin{equation}
ds^2\, =\, d\xi^2\, -\, a^2(\xi) dt^2\,.
\end{equation}
We now define the so-called ``tortoise'' coordinate
\begin{equation}
z(\xi) = \int_{\xi}^\infty \frac{d\zeta}{a(\zeta)}\,. \label{tortoise}
\end{equation} 
The conformal boundary $(\xi \to \infty)$ is at $z=0$, and when $\xi$ is in region I, i.e. $\xi \in \mathbb{R}^+$ then $z$ is real.    
\begin{figure}
\centering
\includegraphics[width=2in]{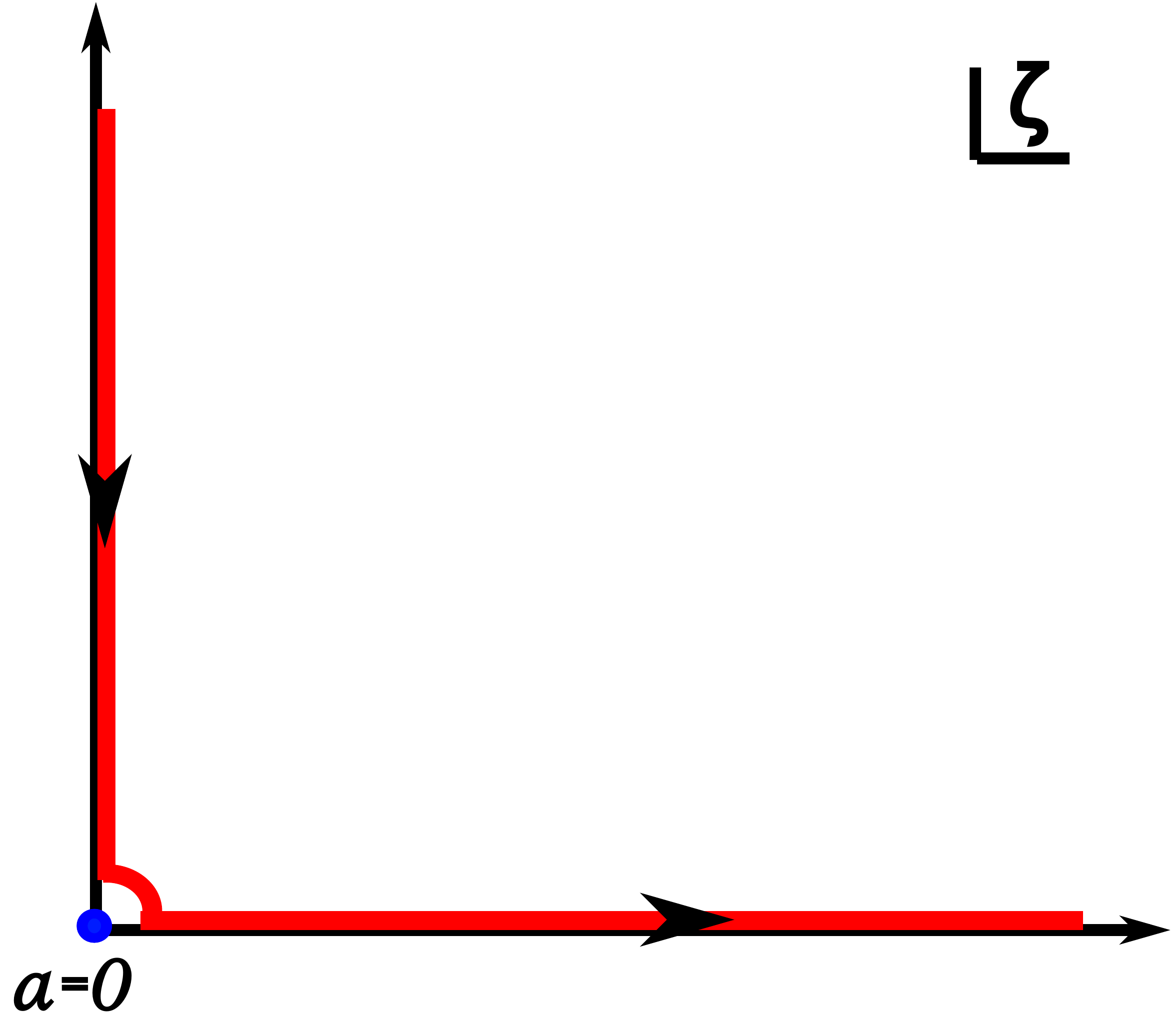}
\caption{\small{The integration contour used to define the tortoise coordinate, covering regions I and II.}}
\label{Lshape}
\end{figure}

When $\xi$ is in region II, i.e. $\xi = i\sigma$ for $\sigma \in \mathbb{R}^+$, the integration contour 
 in \eqref{tortoise} assumes the L-shape indicated in figure \ref{Lshape} in the complex $\zeta$-plane. The contour starts at $\xi=i\sigma$, comes down the imaginary axis, turns the corner at $\zeta=0$ and goes out to $\infty$ along the real axis.  Since we are considering regular solutions, the scale factor near the origin takes the form $a(\zeta)=\zeta+\mathcal{O}(\zeta^3)$ and hence the integrand has a simple pole at $\zeta=0$. Cutting the corner around the pole in a clockwise sense, $z(\zeta)$ acquires an imaginary part  equal to $-i\pi/2$ (one quarter of $(-2\pi i)$ times the residue).   Thus in region II, the tortoise co-ordinate  $z(\zeta)$ has a constant imaginary part $-i\pi/2$.   

In the undeformed AdS spacetime the scale factor $a(\xi) = \sinh{\xi}$ and the tortoise coordinate 
\begin{equation}
z =- \ln{\tanh{\frac{\xi}{2}}}\,,
\end{equation}
which can be analytically continued to region II using $\xi = i\sigma$:
\begin{equation}
z =- \ln{\tan{\frac{\sigma}{2}}} - \frac{i\pi}{2}\,.
\end{equation}

\begin{table}[ht] 
\centering 
\begin{tabular}{| c|c |c |} 
\hline
Section of AdS & $\quad\xi\quad$ & $\quad z\quad$  \\ 
[1ex]	
\hline\hline 
Boundary& $\infty$ & 0\\
[1ex]	
\hline 
Horizon from region I& $0^+$ & $\infty$\\
[1ex]	
\hline 
Horizon from region II& $i\,0^+$& $\infty -\frac{i\pi}{2}$\\
[1ex]	
\hline 
Maximal scale factor
%\footnote{This would be the location of the orbifold singularity of the BTZ black hole for which the radial two dimensional metric looks the same as the $t-\xi$ part of our AdS metric. For AdS it simply marks the maximum of the scale factor in the FRW patch beyond which the space-time continues, the scale factor decreasing until it reaches zero at $\sigma=\pi$ which is just a coordinate singularity through which we can continue into another patch similar to region I.}
& $\frac{i\pi}{2}$ & $-\frac{i\pi}{2}$\\ 
 [1ex] 
\hline 
``Crunch'' & $i\pi$ & $-\infty-\frac{i\pi}{2}$\\
[1ex]	
\hline 
\end{tabular} 
\caption{\small Important sections of pure AdS (with de Sitter slices) in terms of $\xi$ and $z$ (tortoise) coordinates.}
\label{table:ads} 
\end{table}
In tables \ref{table:ads} and \ref{table:deformed} we list the important sections of the space-time, for undeformed AdS and the deformed geometries respectively.  The locations of the conformal boundary and the horizon can be fixed at the same coordinate values for both AdS and the deformed geometries. Given this, the location of maximal scale factor in the FRW region and the location of the crunch $\tilde a =0$, are both shifted by the deformation. Note that for pure AdS, the ``crunch'' at $\xi=i\pi$ is just a coordinate singularity. By the Abbott-Coleman argument, any deformation will lead to a curvature singularity at $\xi = i\sigma_c$ with $\sigma_c < \pi$.
\begin{table}[ht] 
\centering 
\begin{tabular}{| c|c |c |} 
\hline
Section of deformed AdS & $\quad\xi\quad$ & $\quad z\quad$  \\ 
[1ex]	
\hline\hline 
Boundary& $\infty$ & 0\\
[1ex]	
\hline 
Horizon from region I& $0^+$ & $\infty$\\
[1ex]	
\hline 
Horizon from region II& $i\,0^+$& $\infty -\frac{i\pi}{2}$\\
[1ex]	
\hline 
Maximal scale factor& ${i \sigma_m}$ & $w_m\,-\,\frac{i\pi}{2}\,,\quad w_m\in{\mathbb R}$\\ 
 [1ex] 
\hline 
Crunch singularity & $i \sigma_c$ &  $ w_c\,-\,\frac{i\pi}{2}\,,\quad w_m\,>\,w_c\in{\mathbb R}$\\
[1ex]	
\hline 
\end{tabular} 
\caption{\small Sections of deformed AdS  in terms of $\xi$ and $z$ (tortoise) coordinates. On general grounds $\sigma_c < \pi$ and $w_m > w_c$ since the maximal scale factor must be attained {\em before} the crunch singularity.}
\label{table:deformed} 
\end{table}
We now proceed to formulate the Kruskal-like extension, by first defining the retarded and advanced null coordinates as
\bea
u&\,=\,t+z\,,\\\nonumber
v&\,=\,t-z\,,
\eea
in terms of which the metric (suppressing the angular coordinates) takes the form
\begin{equation}
ds^2 \,=\,-a(\xi)^2 du\, dv\,.
\end{equation}
Since the determinant vanishes at the horizon $\xi=0$ this metric is singular there but we can perform a further coordinate transformation to the Kruskal null coordinates 
\begin{align}
U\,=\,-e^{-u}\,,\qquad\qquad V\,=\,e^{v}\,.
\end{align}
These coordinates are defined in region I where $U<0$, $V>0$ but we will show that they can be extended across the horizon.   The metric in the Kruskal coordinates is
\begin{equation}
ds^2 \,=\,-a(\xi)^2 e^{2 z} dU\, dV \,.
\end{equation}
The logarithmically divergent real part of $z$ near $\xi=0$ exactly cancels the zero of $a$ and the metric remains regular as we cross $\xi=0$.    This implies that we can extend it from $U<0$, $V>0$ to $U,V \in (-\infty,\infty)$.  (Recall that, like the tortoise coordinate $z$, $t$ also acquires an imaginary part in region II, as $t \to \chi-i\pi/2$).  In these coordinates curves of constant $\xi$ or equivalently $z$ are hyperbolae,
\begin{equation}
UV \,=\, -e^{-2 z}\,.
\end{equation}
The conformal boundary is given by $UV = -1$.    The horizon is described by the null hypersurfaces $UV=0$.   The turnaround of the scale factor corresponds to $UV=e^{-2 w_m}$ where $w_m=0$ for the undeformed case, while the crunch corresponds to $UV = e^{-2 w_c}$ where $w_c\to-\infty$ for the undeformed case.  
   We cut off the two dimensional representation (with the spatial angular directions suppressed) at the symmetry axis $U=V$, and rotate it about the axis to generate the full higher dimensional Penrose diagram.   

\subsection{Outwardly curved spacelike singularity}
\label{sec:outward}
In order to determine whether the spacelike crunch singularity curves inward or outwards in figure \ref{fig:pda}, we need to first ``straighten" the boundary. This is easily achieved by using an angular parametrisation   $(U,V)\rightarrow (\tau -\psi,\,\tau+\psi)$ defined as,
\begin{align}
U \,=\, \tan{\frac{\tau-\psi}{2}}\,,\qquad\qquad
V \,=\,\tan{\frac{\tau+\psi}{2}} \,.\label{straighten}
\end{align}
The conformal boundary is now represented by the straight vertical lines $\psi = \pm \pi/2$, and the horizon is represented by the null lines $\tau\pm\psi=0$.   In undeformed AdS  the turnaround of the scale factor maps to the lines $\tau = \pm \pi/2$, and the ``crunch'' maps to the null lines $\tau \pm \psi = \pi$.   This takes us back to the Penrose diagram of AdS in figure \ref{fig:pda}\footnote{This set of transformations $(t,\xi)\rightarrow (u,v) \rightarrow (U,V)\rightarrow (\tau,\psi)$ has unsurprisingly taken us back to the AdS metric written in a form conformal to the Einstein static universe, quoted in eq.\eqref{eqn:adsesu}.}.   In the deformed geometry, the turnaround of the FRW scale factor and the crunch are described in the $(\psi,\tau)$ plane by the curve
\begin{equation}
\tan{\frac{\tau-\psi}{2}}\,\tan{\frac{\tau+\psi}{2}}\, =\, e^{-2 w}\,,
\end{equation}
where the maximal scale factor occurs at $w=w_m$ and the crunch at $w=w_c$.  Equivalently, the equation can be rewritten as 
\begin{equation}
\cos{\tau} \,=\, \tanh{w}\,\cos{\psi}\,.
\end{equation}
For fixed $w$, this is a curve anchored between the points $(\psi,\tau)\,=\,\left(-\frac{\pi}{2},\,\frac{\pi}{2}\right)$ and $(\psi,\tau)\,=\,\left(\frac{\pi}{2},\,\frac{\pi}{2}\right)$.   If $w>0$ it lies below the horizontal $\tau=\pi/2$ and curves inward whereas if $w<0$ it lies above it, curving outward.   Showing that the crunch singularity for deformed AdS is outwardly curved boils down to arguing that the crunch is located at $w_c < 0$.  In the limiting case where there is no deformation present, the so-called  ``crunch'' in pure AdS is a null surface as explained above and occurs as $w_c\to -\infty$. It is intuitively reasonable to expect that a small, finite deformation will render $w_c$ finite and negative.  We now provide supporting arguments for this and  confirm the expectation in an analytically tractable example that we will investigate in detail.

The value of $w_c$ is computed  by the real part of the integral of the inverse  scale factor over the whole space-time\footnote{Physically this is the coordinate time taken by a light ray to travel from the boundary to the singularity.}  
\begin{equation}\label{eqn:crunch_coord}
z_c = \int_{i\sigma_c}^\infty \frac{d\zeta}{a(\zeta)} \,.
\end{equation} 
Our argument consists of four key ingredients:
\begin{itemize}
\item
{We will assume that the scale factor in the FRW region has a second zero at $\sigma=\sigma_c <\pi$, the first zero being at the origin. The value of $\sigma_c$ turns out to be less than $\pi$ for the specific example we study in this paper, but  also appears to be true more generally\footnote{This assumption is consistent with the fact that in a number of single scalar models $\ddot{\tilde a} < -\tilde a$ in the FRW patch. For AdS, $\ddot {\tilde a} = -\tilde a$ and $\tilde a = \sin\sigma$. For models with more than one scalar this is empirically observed to be the case \cite{paper2}. For single scalar models, the relevant field equation in the FRW patch reads
\be
\frac{\ddot {\tilde a}}{\tilde a}\,=\,-\frac{2\kappa^2}{d(d-1)}\left(\frac{d-1}{2}\dot\Phi^2\,-\, V_{\Phi}\right)\,,\nonumber
\ee
where we  take $V_{\Phi}$ to be  a negative potential with a single  AdS maximum  where  $\frac{2\kappa^2}{d(d-1)} V_{\Phi}=- 1$ and $\dot \Phi=0$. It follows therefore that in the presence of deformations away from the AdS maximum,
\be
\ddot {\tilde a} \,< \, -\tilde a\,.\nonumber
\ee
We have observed empirically that this condition leads to $\sigma_c < \pi$, assuming $\tilde a(\sigma)=\sigma +\ldots$ for $\sigma \ll1$}. The second zero is the location of the crunch. The existence of the crunch singularity at $\sigma_c$ is guaranteed by the Abbott-Coleman argument \cite{cdl2} (see also appendix A of \cite{jensenruback}).}
\item
{The deformation can be taken to be small enough so that a perturbative solution applies {\em almost everywhere}: The deviation of the scale factor from AdS is small everywhere except near the crunch singularity where the perturbation expansion breaks down. This is again guaranteed on general grounds since the deformed geometries are  obtained by analytic continuation of a (relevant) deformation of EAdS$_{d+1}$ where the deformation can be made parametrically small everywhere in the Euclidean (exterior) region.}
\item
{Close to the singularity as $\sigma\to\sigma_c$ 
%the spatial curvature term in the equations is negligible and 
the scale factor has the asymptotic power law behaviour $\tilde{a} = \tilde{a}_0 (\sigma_c-\sigma)^{\gamma}$ where $0<\gamma \leq 1$. This is true for the analytically solvable example we study below in this paper, and appears to be true more generally, as will be shown in detail elsewhere \cite{paper2}.}
\item
{Breaking up the $z_c$ integral into an AdS-like piece and a near-crunch portion, we can then show that  $z_c$ has a negative and finite real part i.e. $z_c = w_c -i\pi/2$ where $-\infty<w_c  < 0$.}
\end{itemize}

The integral (\ref{eqn:crunch_coord}) for the coordinate of the crunch $z_c$   can be evaluated by breaking it up into three pieces:
\begin{enumerate}
\item{
Near-crunch portion $\sigma_c-\delta <\sigma < \sigma_c$ where $0<\delta \ll 1$ is the width of this asymptotic region.    Here $\tilde{a} = \tilde{a}_0 (\sigma_c-\sigma)^\gamma$.}
\item{
A transition region $\sigma_c-\delta-\epsilon <\sigma < \sigma_c-\delta$ where $0<\epsilon \ll 1$ is the width of this region.}
\item{
The AdS-like part $0<\sigma<\sigma_c-\delta-\epsilon$ in the FRW region, and continuing ($\xi=i\sigma$) through the horizon into the exterior region $0<\xi<\infty$.   For small deformations the scale factor here is well approximated by the AdS form $a(\xi)\,=\,\sinh{\xi}$. }  
\end{enumerate}
Therefore, for parametrically small deformations, and assuming the exponent $\gamma < 1$ we obtain
\bea
z_c &&\,=\, \int_{\sigma_c}^{\sigma_c - \delta}  \frac{d\sigma}{\tilde{a}(\sigma)}  +  \int_{\sigma_c-\delta}^{\sigma_c - \delta-\epsilon}  \frac{d\sigma}{\tilde{a}(\sigma)} +  \int_{i(\sigma_c-\delta-\epsilon)}^{\infty}  \frac{d\xi}{a(\xi)}
\\\nonumber\\\nonumber
&&\,=\, \frac{-\delta^{1-\gamma}}{(1-\gamma)\tilde{a}_0} \, +\,\ln{\frac{\delta'+\delta+\epsilon}{2}}\,+\, \int_{\sigma_c-\delta}^{\sigma_c - \delta-\epsilon}  \frac{d\sigma}{\tilde{a}(\sigma)}\,+\,\mathcal{O}\left(\delta'+\delta+\epsilon\right)^2 \,-\,\frac{i\pi}{2}\,,
\eea
where $\sigma_c = \pi-\delta'$ with $0<\delta' \ll 1$  and we have kept the leading term for $0<\left(\delta'+\delta+\epsilon\right)\ll 1$.  While the transition region can be parametrically small, the contribution from the logarithm dominates the real part of $z_c$, which will thus be negative and finite as required. This completes the demonstration that for (small) deformations of AdS of the type being considered the singularity is curved outwards with respect to the horizontal as indicated in the Penrose diagram we have drawn in figure \ref{fig:pda}.

\section{Spacelike geodesics}
The primary goal of this paper is to examine correlators of the holographically dual QFT on the conformal boundary of region I which is exterior to the bulk horizon. Specifically, we would like to identify probes of the FRW region (region II). 
Below we calculate such correlators in the geodesic limit, which corresponds to operators of sufficiently high dimension, when a WKB-like approximation becomes applicable. We then attempt to interpret the implications of the robust features of these for the appropriate frequency space correlation functions of the boundary QFT. The conformal case, where the bulk is simply AdS$_{d+1}$ spacetime with dS$_d$ slices, is useful for developing some intuition although the geometry is completely smooth (and the ``crunch'' is non-singular). We will then apply the ideas to a specific non-conformal, analytically tractable example with $d=3$.

\subsection{Spacelike geodesics and the maximal expansion slice}
\label{sec:generalities}
We  examine geodesics that penetrate the  horizon at 
$\xi=0$ and thus enter into region II of the bulk geometry, 
and which compute geodesic limits of correlation functions between spatially antipodal points on the boundary. Given the metric for an asymptotically AdS$_{d+1}$ geometry with (global) de Sitter slices,
\be
ds^2\,=\,d\xi^2\,+\, a^2(\xi)\left(-dt^2 \,+\,\cosh^2t\,d\Omega_{d-1}^2\right)\,,
\ee 
in the exterior patch ($\xi^2 >0$), we can determine geodesic paths by extremizing the action
\be
S\,=\,-M\int_{\lambda_i}^{\lambda_f} d\lambda\sqrt{\left(-a^2(\xi)\,\dot t^2\,+\,\dot\xi^2 +a^2(\xi)\cosh^2t\,\dot\phi^2\right)}\,.\label{action}
\ee
Here $\lambda$ is an affine parameter and $M$ denotes the (large) mass of a bulk field which  yields the correlator of a corresponding high dimension operator in the boundary gauge theory.
The end-points of the geodesic are labelled $\lambda_i$ and $\lambda_f$.  We have also  allowed for the trajectory to have angular dependence -- without loss of generality we take this to be along the polar angle $\phi$ of the spatial sphere $S^{d-1}$. The geodesic has a conserved angular momentum, conjugate to the angular coordinate $\phi$:
\be
L\,=\,a^2(\xi)\,\cosh^2 t\,\dot\phi\,.
\ee
The equations of motion automatically imply, as a consequence of reparametrization invariance, the constraint (for spacelike geodesics),  
\be
-a^2(\xi)\,\dot t^2\,+\,\dot\xi^2\, +\,a^2(\xi)\cosh^2t\,\,\dot\phi^2\,=\,1\,.\label{constraint}
\ee
When the angular momentum $L$ vanishes, the geodesics are radially directed, and there is a natural conserved ``energy"\footnote{Real values of ${\cal E}$ will turn out to correspond to imaginary frequencies in the boundary QFT.},
\be
{\cal E}\,=\,a^2(\xi)\,\dot t\,.\label{Econserved}
\ee
This equation, in conjunction with the first order constraint when $L=0$,
\be
\dot\xi^2 \,-\,\frac{{\cal E}^2}{a(\xi)^2}\,=\,1\,,\label{firstorder}
\ee 
determines $t(\lambda)$ and $\xi(\lambda)$, and the trajectory of the radial geodesic. Equivalently, we may directly obtain $t(\xi)$ as a solution of
the condition
\be
\left(\frac{dt}{d\xi}\right)^2\,=\,\frac{{\cal E}^2}{a^2(\xi)\,\left({\cal E}^2\,+\,a^2(\xi)\right)}\,.\label{txifirstorder}
\ee
In certain situations, the second order form of the equation of motion (along with the constraint \eqref{constraint}) may prove to be more convenient,
\be
\ddot\xi \,+\,{\cal E}^2\,\frac{a'(\xi)}{a^2(\xi)}\,=\,0\,.
\label{secondorder}
\ee
This holds for $L=0$ and ${\cal E}\,=\,a^2\,\dot t$\,.

We note that the radial geodesics ($L=0$) naturally connect antipodal points of the spatial sections $\simeq S^{d-1}$ of global de Sitter spacetime on the boundary. While $\dot \phi =0$ for such trajectories, precisely when $a(\xi)=0$, the polar angle $\phi$ can be consistently flipped  to $\pi -\phi$, maintaining the requirement of $L=0$.
We first focus attention on two special solutions which are the easiest to identify.

\paragraph{ Zero-energy geodesics:} These are solutions with $\dot t=0$, passing right through $a(\xi)=0$ which is the origin of the Euclidean region (region I). Explicitly, we have
\be
\xi(\lambda)\,=\,|\lambda-\lambda_0|\,,\qquad
t\,=\,{\rm sgn}(\lambda-\lambda_0)\,t_0\,,\qquad {\cal E}\,=\,0\,,
\ee
where we have taken $\xi=0$ to correspond to the Euclidean origin with $a(\xi=0)\,=\,0$.  
For a smooth geodesic connecting antipodal points we need to append to this the trivial solution for the angular coordinate,
\be
\phi(\lambda)\,=\,\phi_0\,\theta(\lambda_0-\lambda)\,+\,(\pi-\phi_0)
\,\theta(\lambda-\lambda_0)\,.
\ee 

\paragraph{ Maximal expansion FRW slice:} This is a constant-$\xi$ solution to the equation of motion \eqref{secondorder} with 
\be
\xi\,=\,\xi_0\,\qquad a'(\xi_0)\,=\,0\,,\qquad {\cal E}^2
\,=\,-a^2(\xi_0)\,.
\ee
In region I, where $a^2(\xi) >0$, the scale factor $a(\xi)$ is monotonic in general. However, as reviewed above, in the FRW region obtained by analytic continuation ($\sigma \to i\sigma,\, a\to i\tilde a$), the scale factor $\tilde a$ always attains a maximal value prior to the eventual crunch. This  is an immediate consequence of the fact that $\tilde a$ vanishes both at $\sigma=0$ and at the crunch, when  $\sigma\,=\,\sigma_c$. Therefore the constant-$\xi$ geodesic is precisely located on the slice of maximal FRW expansion
\be
\tilde a'(\sigma_m)\,=0\,,\qquad \tilde a_{\rm max}\,\equiv\,\tilde a(\sigma_m)\,, \qquad {\cal E}^2\,=\, +\,\tilde a^2_{\rm max}\,. 
\ee
Therefore the energy variable ${\cal E}$ for this solution is {\em real} and is given by the maximum value attained by the scale factor in the FRW region. Finally, the geodesic extends along the radial direction $\chi$ (see \eqref{region1})
of the spatial hyperbolic slices $\dot t\,=\,\dot \chi\,=\,-1$. Although the maximal slice is actually hidden behind the bulk horizon, it has an important role to play for finite energy radial geodesics joining antipodal points on the conformal boundary of the exterior region.

It turns out that the two special solutions above describe limiting behaviours of generic spacelike, radial geodesics connecting spatially antipodal points on the (global) de Sitter conformal boundary. As we will demonstrate using specific examples, there are two categories of radial solutions with real energy ${\cal E}$: those that have two end-points on the conformal boundary and those with only  one boundary end-point which fall into the FRW crunch singularity. The former have a turn-around point with $\dot \xi =0$. The first order equation \eqref{firstorder} tells us that this happens at some value of the radial coordinate $\xi =\xi_r$ such that
\be
a^2(\xi)\,=\,-{\cal E}^2\,.
\ee
For real ${\cal E}$, this equation is satisfied when the turning point lies behind the bulk horizon in the FRW region. In the exterior (Euclidean) region or region I, the turn-around condition is satisfied only if ${\cal E}$ is purely imaginary. We will see below that real values of ${\cal E}$ correspond to imaginary frequencies $({\cal E}\sim i\omega)$ in boundary correlators.

In the analytically continued coordinates appropriate for region II, 
the geodesic turns around at $\xi_r\,=\,i\sigma_r$ with
\be
\tilde a^2(\sigma_r)\,=\,{\cal E}^2\,.
\ee
For small positive ${\cal E}^2$, the turning point lies just behind the horizon at 
\be
{\cal E}^2\,=\,\tilde a^2(\sigma_r)\approx \sigma_r^2\,,
\ee
and the geodesic is close to the zero energy solution with constant $t$. The origin of the exterior region (region I) is smooth and the scale factor $a^2(\xi)\simeq \xi^2$ for small enough $\xi$. Analytic continuation past the horizon then yields $\tilde a^2(\sigma)\simeq \sigma^2$ for sufficiently small $\sigma$. As the energy is increased, the turning point approaches the maximum of the scale factor. Geodesics with ${\cal E}\lesssim \tilde a_{\rm max}$ remain close to the maximal expansion slice in the FRW region. For energies bigger than this critical value ${\cal E}^2 > \tilde a^2_{\rm max}$, there is no turning point and the geodesics fall into the crunch.

To relate the geodesic to correlators in the boundary QFT\footnote{Note that holographic correlation functions for dS-sliced asymptotically AdS backgrounds can be defined and computed rigorously using methods of holographic RG adapted to these slicings \cite{Papadimitriou:2004ap, deHaro:2000vlm, Bianchi:2001kw, kw}.}, we need to calculate the action for the geodesic solution, and further be able to relate the conserved `energy' to a frequency-like observable in the boundary theory. Before studying the deformed AdS geometries, we first focus attention on the pure AdS theory.

\subsection{Spacelike geodesics in pure AdS}

Spacelike geodesics connecting two spacelike separated points on the boundary can be studied in the de Sitter slicing of pure AdS spacetime, by taking $a(\xi)=\sinh\xi$. 

\subsubsection{Geodesics with ${\cal E}<1$}
Then the solution to the first order constraint \eqref{firstorder}, for a fixed energy ${\cal E}$, is 
\be
\cosh\xi\,=\,\sqrt{1-{\cal E}^2}\,\cosh(\lambda-\lambda_0)\,,\qquad {{\cal E} <1}\,.
\label{xilambda}
\ee
Here $\lambda$ is the affine parameter along the geodesic and $\lambda_0$, an arbitrary real constant which can be set to zero without loss of generality. This solution assumes that ${\cal E}<1$, from which  it immediately follows that there are two values of $\lambda$ corresponding to $\xi=0$ which is the horizon separating regions I and II. Of the two values of $\lambda$, the one with $\lambda < \lambda_0$ yields the point of entry into the horizon (when $\dot \xi <0$) and the other corresponds to the exit point ($\dot \xi>0$) of the geodesic from  the horizon.  

The global de Sitter time coordinate $t$ can also be solved for by combining eqs.\eqref{Econserved} and \eqref{firstorder}
\be
\frac{dt}{d\xi}\,=\,\pm \frac{{\cal E}}{\sinh\xi\sqrt{{\cal E}^2+\sinh^2\xi}}\,,
\ee
which can be integrated to yield
\be
t\,=\,t_{1(2)}\,+\,\tanh^{-1}\left(\frac{{\cal E}\,\cosh{\xi}}{\sqrt{\sinh^2\xi\,+\,{\cal E}^2}}\right)\,-\,\tanh^{-1}{\cal E}\,,
\ee
in the exterior region picking the solution branch with $\frac{d\xi}{dt}<0$. The constants of integration $t_1$ and $t_2$ are the values of the time coordinate at the two end-points of the geodesic on the conformal boundary at $\xi\to\infty$.  At its `starting point' with $\lambda\to -\infty$ we take $t=t_1$.

Upon entering region II behind the horizon as $\lambda$ is increased (in coordinate time this occurs asymptotically as $t\to \infty$), we can obtain the geodesic in the FRW region by the straightforward analytic continuation $\xi\to i\sigma$ so that,
\be
t\,=\,-\frac{i\pi}{2}\,+\,t_1\,+\,\coth^{-1}\left(\frac{{\cal E}\,\cos{\sigma}}{\sqrt{{\cal E}^2\,-\,\sin^2\sigma}}\right)\,-\,\tanh^{-1}{\cal E}\,.\label{part1frw}
\ee
The imaginary part here can also be understood in terms of the phase picked up whilst traversing the L-shaped contour and avoiding $\xi=0$ as shown in 
figure \ref{Lshape}.
In the FRW patch the spatial slices are hyperboloids with  radial coordinate $\chi\,=\,t\,+\,i\pi/2$ as in eq.\eqref{toFRW}. We note that since $\chi$ is a radial coordinate, it cannot be negative. In fact, when the geodesic passes through the spatial origin at $\chi=0$, we are required to change the solution branch.

Assuming that the FRW patch solution \eqref{part1frw} turns around before getting to the origin (${\rm Re}(t)=\chi=0$), the turnaround point is reached 
 when the affine parameter $\lambda=\lambda_0$ in eq.\eqref{xilambda}, so that
\be
\sigma\,=\,\sigma_{\rm turn}\,=\,\sin^{-1}{\cal E}\,, \qquad {\cal E}<1\,,
\ee
and
\be
t\,=\,t_{\rm turn}\,=\,-\frac{i\pi}{2}\,+\,t_1\,-\,\tanh^{-1}{\cal E}\,.
\ee
Since the solution \eqref{part1frw} turns around {\em before} $\chi$ reaches zero,  we must require $t_1 > \tanh^{\rm -1}{\cal E}$.  Past this point, the FRW time $\sigma$ along the geodesic and the radial coordinate $\chi$ simultaneously decrease so that we pick the branch of the solution that has $\frac{d\sigma}{d\chi}>0$:
\be
t\,=\,-\frac{i\pi}{2}\,+\,t_1\,-\,\coth^{-1}\left(\frac{{\cal E}\,\cos{\sigma}}{\sqrt{{\cal E}^2\,-\,\sin^2\sigma}}\right)\,-\,\tanh^{-1}{\cal E}\,.
\ee
This branch of the solution is valid for $0\leq \chi\leq {\rm Re}(t_{\rm turn})$.
After the geodesic reaches $\chi=0$, the solution takes the form,
\be
t\,=\,-\frac{i\pi}{2}\,+\,t_2\,+\,\coth^{-1}\left(\frac{{\cal E}\,\cos{\sigma}}{\sqrt{{\cal E}^2\,-\,\sin^2\sigma}}\right)\,-\,\tanh^{-1}{\cal E}\,,\label{part2frw}
\ee 
which applies in the range $0\leq \chi <\infty$ i.e. between the origin of the hyperbolic slice and the second horizon crossing. The constant of integration $t_2$ is related to $t_1$ and determined by matching the two solutions at $\chi={\rm Re}(t)=0$:
\be
t_1\,+\,t_2\,=\,\,+\,2\tanh^{-1}{\cal E}\,.\label{meanE}
\ee
After the second horizon crossing the `outgoing geodesic' is described by the continuation of  \eqref{part2frw} to the exterior region,
\be
t\,=\,\,t_2\,+\,\tanh^{-1}\left(\frac{{\cal E}\,\cosh{\xi}}{\sqrt{{\cal E}^2\,+\,\sinh^2\xi}}\right)\,-\,\tanh^{-1}{\cal E}\,.
\ee
The geodesic now approaches the conformal boundary $\xi\to\infty$ as $t\to t_2$.

We have completely characterised the geodesic joining two spatially antipodal points (with polar angles $\phi$ and $\pi-\phi$) at (global) de Sitter times $t_1$ and $t_2$ on the conformal boundary of the pure AdS geometry. Note that, given two such points, the energy ${\cal E}$ is fixed by the mean value of the two times $t_1$ and $t_2$. 

\begin{figure}
\centering
\includegraphics[width=2.7in]{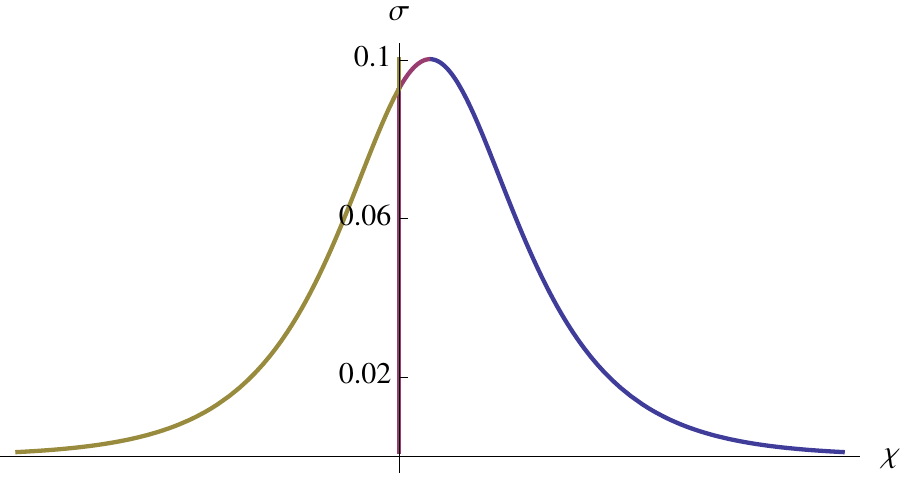}\hspace{0.4in}
\includegraphics[width=2.7in]{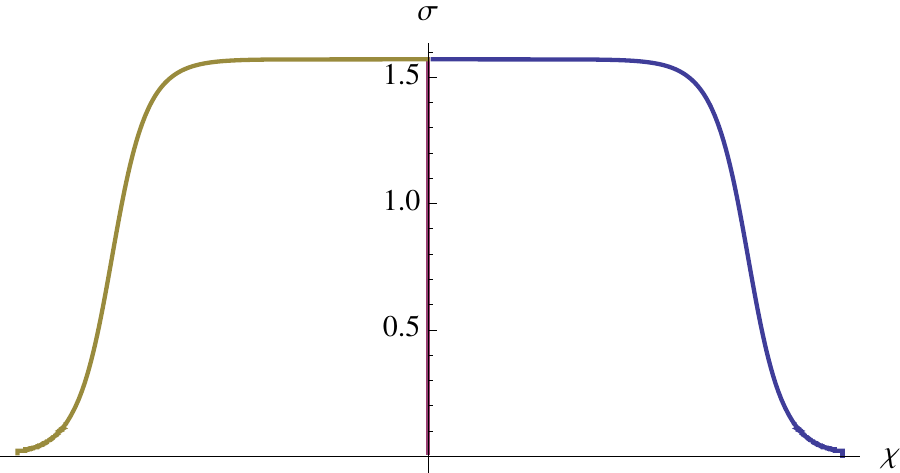}
\caption{\small{{\bf Left:} The continuation of the spacelike geodesic into the FRW patch of pure AdS spacetime with ${\cal E}=0.1$ and $t_1=0.5$. {\bf Right:} For energies ${\cal E}\simeq 1$, the geodesic remains close to the slice of maximum expansion $\sigma =\frac{\pi}{2}$ in the FRW patch. As the radial geodesics pass through the origin ${\chi=0}$ their angular orientation changes from a given polar angle $\phi$ (portion in blue) to $(\pi -\phi)$ (red portion), thus connecting antipodal points on the conformal boundary.   }}
\label{fig:adsfrw}
\end{figure}

Figure \ref{fig:adsfrw} shows the behaviour of the spacelike geodesic in the FRW patch. In particular as the energy ${\cal E}$ approaches the maximum value of the scale factor, which is $\tilde a_{\rm max}=1$ for pure AdS spacetime, the geodesic hugs the maximal expansion slice.
An important corollary of this is that for late (de Sitter global) times i.e. as $t_{1,2}\to\infty$, the geodesic energy ${\cal E}$ also approaches unity and therefore the solution is well approximated by the slice of maximal expansion behind the horizon. We will see that the behaviour persists for the deformed AdS geometries as well.

\subsubsection{Geodesic action for ${\cal E}<1$}
The action for the geodesic computes a two-point correlator of some operator with  large conformal dimension $\Delta$. In the geodesic limit $\Delta \gg 1$, the corresponding  bulk field will have a large mass $M \simeq \Delta\gg 1$ (in units where the AdS radius is set to one). The geodesic length is obtained by integrating the regulated action along the L-shaped contour as the geodesic enters and subsequently exits the horizon, effectively retracing the contour. In the process the imaginary  piece cancels and we find
\bea
S&&=\,2M\int_{i\sigma_{\rm turn}}^{\xi_\infty}d\xi\,\left({1+\frac{{\cal E}}{\sinh^2\xi}}\right)^{-1/2}\,\\\nonumber\\\nonumber
&&=\,M\left[2\xi_{\infty} \,-\,\ln\left(1-{\cal E}^2\right)\right]\,.
\eea
The integral is formally divergent and we have regulated it by introducing a (UV) cutoff $\xi_\infty$. The divergent piece can be subtracted away unambiguously, up to finite terms that affect the overall normalisation of the correlator in question. We define the regulated action as
\be
S_{\rm reg}\,\equiv\,S\,-\,M(2\xi_{\infty}\,-\,2\ln 2)\,.
\ee
The two-point function of the dual operator ${\cal O}_\Delta$ (with $\Delta\simeq M$) is therefore
\be
\langle{\cal O}_\Delta(t_1,\, \phi)\, {\cal O}_\Delta(t_2,\, \pi-\phi)\rangle\,=\,e^{-S_{\rm reg}}\,=\,\left[4\,\cosh^{2}\left(\frac{t_1+t_2}{2}\right)\right]^{-M}\,.
\label{ads2pt}
\ee
Here we have used the relation \eqref{meanE} between the geodesic energy and the temporal coordinates of the geodesic end-points. As we explain below, this result can be deduced simply as a consequence of conformal invariance. Prior to this, we examine the fate of geodesics with ${\cal E}>1$.

\subsubsection{Geodesics with ${\cal E}>1$}
For energies bigger than the threshold value, ${\cal E}>1$, the first order equation \eqref{firstorder} is solved by 
\be
\cosh\xi\,=\,\sqrt{{\cal E}^2-1}\,\sinh\left|\lambda_0-\lambda\right|\,,
\ee
in the exterior region $\xi^2 >0$. Unlike the situation with ${\cal E} <1$, a solution that enters the horizon does not turn around and exit. Instead, it reaches the point $\sigma =\pi$ in the FRW patch. The temporal coordinate in the exterior region is given by
\be
t\,=\, t_1\,+\,\coth^{-1}\left(\frac{{\cal E}\,\cosh \xi}{\sqrt{{\cal E}^2\,+\,\sinh^2\xi}}\right)\,-\,\coth^{-1}{\cal E}\,,
\ee
which, after horizon crossing becomes
\be
t\,=\, t_1\,-\,\frac{i\pi}{2}\,+\,\tanh^{-1}\left(\frac{{\cal E}\,\cos \sigma}{\sqrt{{\cal E}^2\,-\,\sin^2\sigma}}\right)\,-\,\coth^{-1}{\cal E}\,.
\ee
The geodesic has no turning point. It reaches the spatial origin $\chi= {\rm Re}(t)=0$, beyond which the solution continues  (with a sign change  since $\chi$ is a nonnegative radial coordinate) onwards until it reaches the null surface at $\sigma =\pi$, never exiting region II.

\subsubsection{Map from global to  dS-slicing of AdS}
\paragraph{Mapping correlators:} De Sitter spacetime in $d$ dimensions is conformal to the Einstein static universe (ESU) or the flat cylinder ${\mathbb R}\times S^{d-1}$. The conformal map  is achieved by a simple coordinate transformation on the global de Sitter time
\bea
ds^2\left.\right|_{{\rm dS}_d} &&\,=\,-dt^2\,+\,\cosh^2t\,d\Omega_{d-1}^2\,=\,
\Lambda_{\rm dS}^2(\tau)\,\left(-d\tau^2\,+\,d\Omega_{d-1}\right)\,,\\\nonumber\\\nonumber
&& \cos \tau\,=\,{\rm sech}\, t\,,\qquad\qquad \Lambda_{\rm dS}(\tau)\,=\,\sec \tau\,.
\eea
This maps future and past infinities of global de Sitter time to the finite times $\tau=\pm\frac{\pi}{2}$ on the cylinder. 
Therefore, in a conformal field theory on dS$_d$, the two-point correlator of an operator with conformal dimension $\Delta$ satisfies
\be
\langle {\cal O}_{\Delta}(t_1,\phi_1)\,{\cal O}_{\Delta}(t_2, \phi_2)\rangle_{{\rm dS}_d}\,=\,
\Lambda_{\rm dS}(\tau_1)^{-\Delta}\,\Lambda_{\rm dS}(\tau_2)^{-\Delta}\,\langle {\cal O}_{\Delta}(\tau_1, \phi_2)\,{\cal O}_{\Delta}(\tau_2,\phi_2)\rangle_{\rm ESU}\,,\label{esu2ds}
\ee
where $\phi_{1,2}$ represent the  (spatial) angular coordinates of the insertion points of the two operators on $S^{d-1}$. Conformal invariance determines the form of the two-point correlator on the cylinder ${\mathbb R}\times S^{d-1}$ which is conformal to ${\mathbb R}^d$. For antipodal points $\phi_2\,=\, \pi-\phi_1$, it immediately follows that 
\be
\langle {\cal O}_{\Delta}(\tau_1, \phi_2)\,{\cal O}_{\Delta}(\tau_2,\phi_2)\rangle_{\rm ESU}\,\sim\, \left[1+ \cos(\tau_1-\tau_2)\right]^{-\Delta}\,.\label{2ptesu}
\ee
This is regular for all ${\tau_1}, \tau_2$, except when ${\tau_1-\tau_2}\,=\pi$, which is a lightcone singularity, when the spatially antipodal points are null separated.
Now the correlator in dS$_d$ is given by the conformal transformation \eqref{esu2ds}\,
\be
\langle {\cal O}_{\Delta}(t_1,\phi_1)\,{\cal O}_{\Delta}(t_2, \phi_2)\rangle_{{\rm dS}_d}\,\sim\,\left(\frac{1+ \cos(\tau_1-\tau_2)}{\cos\tau_1\,\cos\tau_2}\right)^{-\Delta}\,=\,\left[2\cosh\left(\frac{t_1+t_2}{2}\right)\right]^{-\Delta}\,.
\ee 
This precisely matches (up to a normalization), the result obtained by the geodesic computation \eqref{ads2pt}. There are two noteworthy points here: First, and this is specific to the undeformed CFT (dual to AdS gravity), the geodesic approximation captures the exact CFT correlator on the de Sitter background. Secondly, for late global times $t_{1,2}\to \infty$ or  equivalently, near the `end of time' in the Einstein static universe $\tau_{1,2}\to \frac{\pi}{2}$, the de Sitter space correlation functions factorize and are vanishing:
\be
\langle {\cal O}_{\Delta}(t_1,\phi_1)\,{\cal O}_{\Delta}(t_2, \phi_2)\rangle_{{\rm dS}_d}\,\sim\, e^{-\Delta(t_1+t_2)/2}\,.
\ee
It turns our that this behaviour persists with certain modifications in the presence of relevant deformations in the CFT. Moving from the CFT on global de Sitter spacetime to the description on the Einstein static universe can be understood in the bulk as a coordinate transformation that takes AdS with dS-slicings to global AdS spacetime.

\paragraph{Geodesics in global AdS:} AdS$_{d+1}$ spacetime in global coordinates can be written in coordinates that are conformal to the Einstein cylinder ${\mathbb R}\times S^d$ (not to be confused with the conformal boundary  ${\mathbb R}\times S^{d-1}$):
\be
ds^2\,=\,\sec^2\psi\left(-d\tau^2\,+\,d\psi^2\,+\,\sin^2\psi\,d\Omega_{d-1}^2\right)\,,\qquad 0 \leq \psi < \frac{\pi}{2}\,.
\ee
These are related to the coordinates in the dS-slicing of AdS through
\be 
\tan\psi\,=\,\sinh\xi \cosh t\,,\qquad\qquad\tan \tau \,=\, \tanh\xi\sinh t\,. \label{coordchange}
\ee
It is a very useful exercise to see how spacelike geodesics in global AdS relate to the ones discussed above in the dS-sliced description. Taking  $\lambda$ as the affine parameter along the geodesic, following standard steps, we obtain two conserved charges
\be
E\,=\,\sec^2\psi\,\dot \tau\,,\qquad\qquad L\,=\,\tan^2\psi\,\dot\phi\,,
\ee
the energy and angular momentum respectively in the global slicing. All derivatives are with respect to the affine parameter. Focussing attention on radial geodesics $(L=0)$, the first order equation for the radial coordinate $\psi$
 becomes
 \be
 \sec^2\psi\,\dot\psi^2\,-\,E^2\,\cos^2\psi\,=\,1\,.
 \ee
 The solution for the radial geodesic with energy $E$ is given by
 \be
 \tan\psi\,=\,\pm\sqrt{1+E^2}\,\sinh(\lambda - \lambda_0)\,,\qquad \tan(\tau-\tau_0)\,=\,E\,\tanh(\lambda -\lambda_0)\,,\label{globalgeo}
 \ee
 which passes through the origin of AdS ($\psi=0$) at $\tau=\tau_0$ and affine parameter $\lambda=\lambda_0$. Eliminating the affine parameter, the solution can be represented compactly as
\be
\sin\psi\,=\,\pm\frac{\sqrt{1+E^2}}{E}\,\sin(\tau-\tau_0)\,.
\label{globalsol}
\ee
This is the solution to the first order differential equation 
\be
\left(\frac{d\tau}{d\psi}\right)\,=\,\pm\frac{E\, \cos\psi}{\sqrt{1+E^2 \cos^2\psi}}\,.\label{taupsigeo}
\ee 
Note that the solution with vanishing global energy $E=0$ is the constant time geodesic $\tau =\tau_0$. On the other hand, a geodesic at constant de Sitter time $t_1$ has de Sitter energy ${\cal E}=0$ whilst its global energy is given by $E=\sinh t_1$ (where we have used 
$\sin\tau\,=\,\tanh t \sin\psi$). The proper length of the geodesic (times the mass) is then
\bea
S_{\rm ESU}&&=\,2M\int_0^{\frac{\pi}{2}-\epsilon}d\psi \,\sec\psi\sqrt{1-\tau_\psi^2}\\\nonumber\\\nonumber
&&=\,M(2\ln 2\,-\,\ln(1+E^2)\,-\,2 \ln\epsilon\,).
\eea
From eq.\eqref{globalgeo}, taking the end-points $(\lambda\to\pm\infty)$ of the geodesic to be at times $\tau_{1}$ and $\tau_2$, we obtain the relation
\be
\tau_2-\tau_1\,=\,2\tan^{-1}E\,.\label{meanE2}
\ee 
The regulated action obtained by subtracting the divergent piece yields the two-point function of the dual CFT operator with conformal dimension $\Delta \simeq M$,
\be
\langle{\cal O}_\Delta(t_1,\, \phi)\,
{\cal O}_\Delta(t_2,\,\pi-\phi) \rangle_{\rm ESU}\,=\,
e^{-S_{\rm ESU,\,reg}}\,=\, \left[4\cos^2\left(\frac{\tau_1-\tau_2}{2}\right)\right]^{-M}\,.
\ee
The result is, of course, guaranteed by conformal invariance. Note that here we set out to compute CFT correlators on ${\mathbb R}\times S^{d-1}$. To directly obtain the same on dS$_d$, but using the global slicing, we would need to employ a cutoff at large fixed $\xi\,=\xi_\infty$. Then the cutoffs on the two end-points of the geodesic $\psi_{1,2}\,=\,\pi/2-\epsilon_{1,2}$ get related to $\xi_\infty$ as
\bea
\epsilon_{1,2}\,\simeq\,e^{-\xi_{\infty}}\,{\rm sech}\, t_{1,2}\,.\label{dscutoff}
\eea  
Plugging this into the geodesic action and subtracting off the piece proportional to $\xi_{\infty}$ yields the correct de Sitter space correlator.

\begin{figure}
\centering
\includegraphics[width=3.5in]
{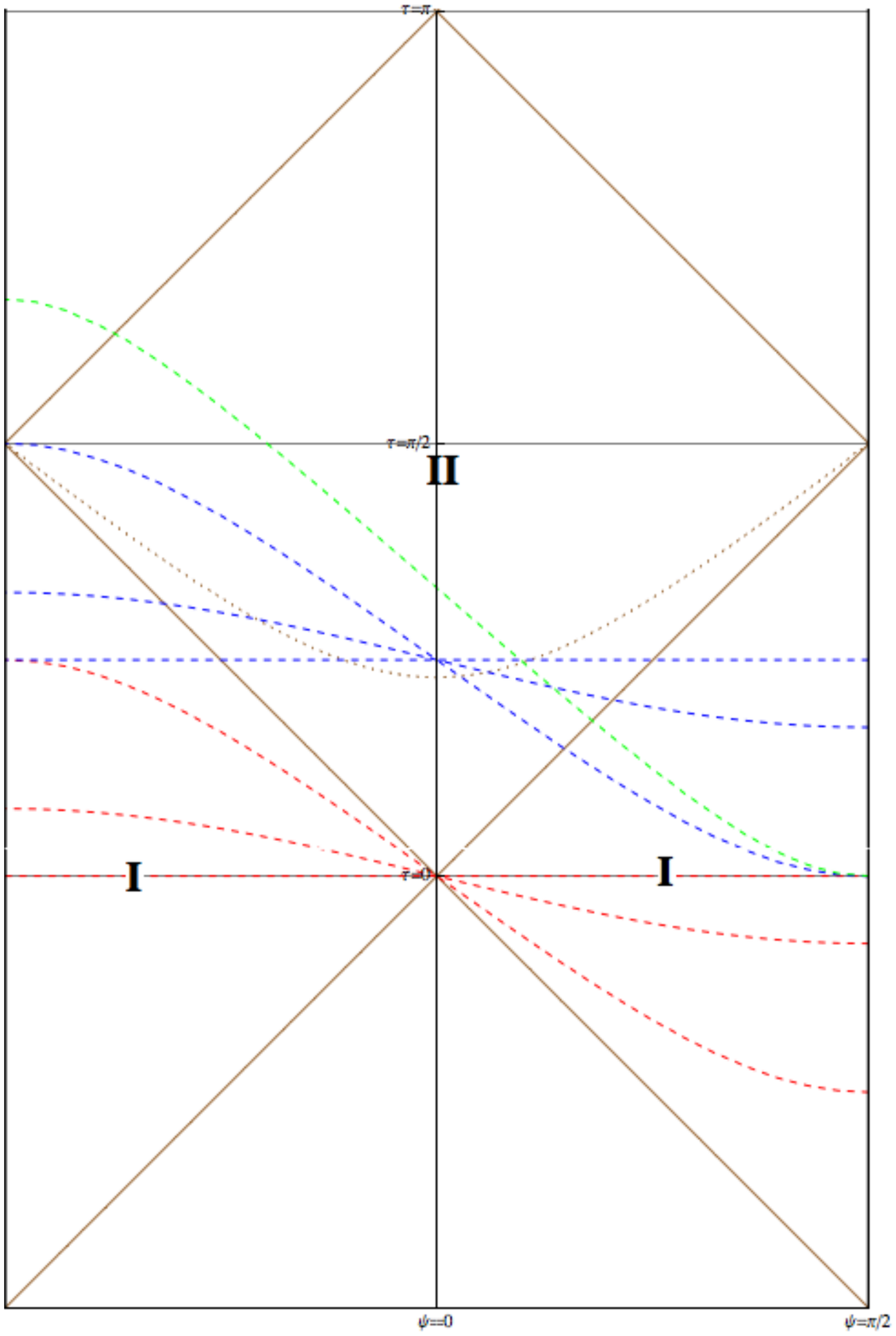}
\caption{\small{Some radial spacelike geodesics in undeformed global AdS.
The $E=0$ geodesic passing through $\tau=0$ coincides with the horizontal red line. The dashed red lines  are geodesics with global energies $E=1/4$ and $E=1$, passing through $\tau_0=0$.  Blue dashed lines show $E=0$, $E=1/4$ and $E=1$ geodesics with $\tau_0=\pi/4$.  The green dashed line shows an $E=\sqrt{3}$ geodesic with $\tau_0=\pi/3$ that does not return to region I.  The dotted line is a contour of constant $\sigma$ in region II.}}
\label{fig:adsgeodesics}
\end{figure}

\paragraph{Relating geodesics in global and dS-slicings:} We can now identify the map between geodesics in global AdS and the dS-sliced geometry via the two corresponding sets of parameters (energy and the initial time):
\bea
{\cal E}\,=\,\sqrt{1+E^2}\,\sin\tau_0\,,\qquad
\qquad\sinh t_1\,=\,\frac{\tan \tau_0 - E}{1+ E\tan \tau_0}\,.
\eea
Here ${\cal E}$ and $t_1$ are the de Sitter energy and initial (de Sitter) time respectively, whilst $E$ and $\tau_0$ are the global energy and global time when the geodesic reaches the origin of global AdS. The above map between the parameters (and its inverse) follow directly from eqs.\eqref{meanE}, \eqref{meanE2}, \eqref{globalsol}  and the transformation \eqref{coordchange} between the two slicings of AdS. 

A geodesic at constant global time $\tau=\tau_0$ has zero global energy $E=0$. It maps to a geodesic in the dS-sliced geometry with 
de Sitter energy ${\cal E}\,=\,\sin\tau_0$ and initial time $\sinh t_1\,=\,\tan\tau_0$. On the other hand, the geodesic with vanishing de Sitter energy ${\cal E}=0$ corresponds to a global AdS geodesic with energy $E\,=\,-\sinh t_1$, passing through $\tau_0\,=\,0$.

Figure \ref{fig:adsgeodesics} shows some geodesics in global AdS. The threshold between geodesics that cross the horizon twice and return to region I, and those that fall into the ``crunch'' (pure AdS is non-singular), is controlled by the de Sitter energy ${\cal E}$. In the limit $E\to \infty$, the global geodesics become null \eqref{globalsol}. As is apparent in figure \ref{fig:adsgeodesics}, irrespective of their global energy $E$, all geodesics passing through
$\tau_0=0$ have both their endpoints in region I. In particular, they have vanishing de Sitter energies and correspond to geodesics that do not penetrate into the FRW region in the dS-sliced description.   When the value of $\tau_0$ is changed, the geodesics may or may not have both end-points in region I. For example, the geodesic with $E\,=\,1/4$ and $\tau_0 \,=\,\pi/4$ is shown dipping in and out of a contour of constant $\sigma$ in region II. There is, however, nothing special about the turnaround value of $\sigma$ in global coordinates.

 \subsection{Geodesics in deformed AdS: a first pass}
 Before turning to a detailed study of spacelike correlators in the geodesic limit in an analytically tractable example, we outline the general strategy for the analysis.  As we have explained in section \ref{sec:intro} the deformed EAdS metric can be put into a form which is conformal to AdS by transforming to the hatted coordinates 
\begin{equation}
\hat{\xi} = 2 \tanh^{-1}{e^{-z(\xi)}}\,,
\end{equation}
where $z$ is the tortoise coordinate
\begin{equation}
z(\xi) = \int_{\xi}^\infty \frac{d\zeta}{a(\zeta)}\,. 
\end{equation} 
The deformed metric can be written in the form, 
\begin{equation}
ds^2 \,=\, \Lambda^2 \left(d\hat{\xi}^2 \,+\, \sinh^2{\hat{\xi}}\,(-dt^2 +\cosh^2{t}\,d\Omega_{d-1}^2)\right)\,,\label{defconformal}
\end{equation}
with the conformal factor given as 
\begin{equation}
\Lambda(\hat{\xi})\, =\, \frac{a\left(\xi(\hat{\xi})\right)}{\sinh{\hat{\xi}}}\, =\, a \sinh{z}\,.
\end{equation}
For pure AdS, $a=\sinh{\xi}$, $z = -\ln{\tanh{\frac{1}{2}\xi}}$ and  $\Lambda=1$.     Since the deformations should be  asymptotically AdS, we require $\Lambda\to 1$ near the conformal boundary.   

To solve for geodesics in the deformed space-time, we could use any of the different coordinate systems employed for undeformed AdS, provided we also transform the conformal factor.  The de-Sitter/FRW coordinates are not regular at the horizon.  It is more convenient  to work in global coordinates, so that
\begin{equation}
ds^2 \,=\, \Lambda^2 \sec^2{\psi}\left(-d\tau^2 + d\psi^2 + \sin^2{\psi}\, d\Omega_{d-1}^2\right)\,,
\end{equation}
where  the conformal factor is non-trivial,
\begin{equation}
\Lambda\,=\,\Lambda\left(\cosh^{-1}{(\sec{\psi}\cos{\tau})}\right)\,,
\end{equation}
and explicitly depends on global time $\tau$.
The $\tau$-dependence of the metric implies that  there is no longer a conserved global energy $E$, except asymptotically, and we do not have a first integral of the equations of motion. For radially directed (spacelike) geodesics we can use the square root action
\begin{equation}\label{eqn:globalsquarerootaction}
S \,=\,M \int_0^{\frac{\pi}{2}-\epsilon} d\psi\, \Lambda \sec{\psi} \sqrt{1-\tau_{\psi}^2}\,. 
\end{equation}   
The second-order equation of motion for the implicit geodesic $\tau(\psi)$ can be integrated numerically from the boundary with the initial conditions
\begin{equation}
%\label{eqn:globalsquarerootactionBC}
\tau(\pi/2-\epsilon)\,=\, \tau_1\,,\qquad\qquad \tau'(\pi/2-\epsilon)\,=\, - E\,\epsilon\,.\label{geodesicbc}
\end{equation}     
These initial conditions follow from AdS asymptotics using equation \eqref{taupsigeo} which relates the gradient $\tau'$ to the asymptotic global energy. As seen earlier for pure AdS geometries, the cutoff $\epsilon$ in global coordinates should be chosen correctly so as to yield the correlator in the dS-sliced deformed geometry. Note also that despite the absence of a conserved energy $E$ in (conformally) global slicing, radial geodesics in the dS-sliced deformed geometry \eqref{defconformal} will have an associated de Sitter energy $\cal E$, as explained in section \ref{sec:generalities}.

\section{Deformed AdS$_4$ example}
\label{sec:yiannis}
A very interesting, analytically tractable example of a deformation of EAdS$_4$ is provided by the supergravity solution discussed in  \cite{yiannis1} (see also \cite{yiannis0, yiannis2}). This arises from a consistent, single scalar truncation of $\mathcal{N}=8$ gauged supergravity in four dimensions with the following action and scalar potential (in  Euclidean signature):
\bea
&& S_{\rm truncated}\,=\,\int d^4x\,\sqrt{g}\,\left[-\frac{1}{2\kappa^2}{R}\,+\,\frac{1}{2}g^{\mu\nu}\partial_\mu\Phi\partial_\nu\Phi\,+\,V_{2/3}(\Phi)\right]\,,\label{23scalarpot}\\\nonumber\\\nonumber
&&V_{2/3}\,=\, -\frac{3}{\kappa^2 L^2_{\rm AdS}} \cosh\left({\sqrt{\frac{2}{3}}\kappa\,\Phi}\right)\,,
\eea
where $\kappa^2 = 8\pi G_4$ is  the four dimensional Newton's constant and the scalar $\Phi$ is {\em minimally} coupled to the curvature. 
The potential is the specialisation to $d=3$ of the so-called ``2/3'' potential of \cite{yiannis1}. It is worth emphasising that this particular truncation and its associated  scalar potential  are distinct from the single scalar truncation    considered in the original works on big crunch duals \cite{HH1, HH2}.
%\subsection{Closed form solution on $S^3$}\label{sec:closedformsolution}  
The scalar potential can be derived from a superpotential
\begin{equation}
W_{2/3}\, =\, -\frac{1}{2\kappa^2 L_{\rm AdS}}\, e^{-\sqrt{\frac{3}{2}}\,\kappa\Phi}\,\left(1\,+\, 3\, e^{2\,\sqrt{\frac{2}{3}}\,\kappa\Phi}\right)\,,\label{23pot}
\end{equation}
assuming the supergravity coupling $g=2$. Interestingly, there exists a field redefinition which maps the above system to a scalar field conformally coupled to Einstein gravity with negative cosmological constant and quartic scalar potential \cite{mtz}. The scalar potential in \eqref{23scalarpot} has an AdS$_4$ maximum at $\Phi=0$ with small perturbations having mass $M^2 L^2_{\rm AdS} \,=\,-2$. This value of the mass-squared lies in the window $-9/4 < M^2 L^2_{\rm AdS} < -5/4$, permitting consistent quantization with two types of boundary conditions 
\cite{bf, kw}. The two boundary conditions, Dirichlet and Neumann, lead to   correspondence with a dual  CFT operator of dimension $\Delta =2$ and $\Delta=1$ respectively.

Choosing units where $\kappa=1$, $L_{\rm AdS}=1$, the above system has a one-parameter family of regular solutions with $S^3$ slices:   
\bea
&&ds^2 \,=\, \left(1-\frac{f(u)^2}{6}\right)\left[\frac{du^2}{u^2(1+u^2)}\,+\,\frac{1}{u^2}\, d\Omega_3^2\right]\,,\label{defads}  \\\nonumber \\\nonumber
&&\Phi \,=\, \sqrt{6}\tanh^{-1}{\left(\tfrac{f}{\sqrt{6}}\right)}\,,\qquad f(u)\,=\, \frac{f_0 \,u}{\sqrt{1+u^2}+u\sqrt{1+f_0^2/6}}\,.
\eea
The constant $f_0$ is related to the value of the field at the Euclidean origin $u\to \infty$, where the metric is effectively flat. The conformal boundary is at $u \to 0$. This background is the  $\mu=0$, $k=1$ and $d=3$ case of the solutions presented in \cite{yiannis1}.  The metric is conformally EAdS$_4$ as is expected for a deformed EAdS$_{d+1}$ geometry with $S^d$ slices\footnote{A  special feature of this solution is that upon performing the field redefinition \cite{yiannis1,mtz} which maps it to the conformal frame, the metric becomes undeformed AdS$_4$.}. In particular, the metric multiplying the conformal factor in eq.\eqref{defads} can be put in the standard AdS form by the coordinate transformation $u^{-1}\,=\,\sinh \xi$. 

Let us now try to understand the behaviour of the scale factor $a(\xi)$ in the Euclidean and FRW regions upon analytic continuation to Lorentzian signature. As  before the $a(\xi)$ is defined via the deformed Euclidean geometry,
\be
ds^2\,=\,d\xi^2\,+\,a(\xi)^2\,d\Omega_3^2\,.
\ee
It is always possible to find the variable change from $\xi$ to $u$, but this is not really necessary at this juncture.
The boundary asymptotics ($u\to 0$) yield
\bea
%&&
\Phi(u)\,  \simeq\, f_0 u - f_0 \sqrt{1+\tfrac{f_0^2}{6}} \,u^2
%\\\nonumber \\\nonumber
%&& 
\,,\qquad\qquad a(\xi(u)) \, \simeq\, \frac{1}{u}\, -\, \frac{f_0^2}{12}\,u\,.\label{asymptriple} 
\eea
%It was argued in \cite{Hertog:2004dr} that for an AdS$_4$ scalar with mass $M^2=-2$, there exists a one parameter family of AdS-invariant quantisations.  This can be understood as follows. 
It is well known, following \cite{kw}, that given the two independent fall-offs
\be
\Phi \,\sim\,\alpha\, u\,+\,\beta \,u^2\,,
\ee
one may interpret $\alpha$ as a deformation by a $\Delta=2$ operator and $\beta$ as the corresponding VEV (Dirichlet boundary conditions). Alternatively, one may adopt  Neumann boundary conditions wherein $\beta$ is viewed as a deformation by a $\Delta=1$ CFT operator and $\alpha$ as its VEV. In addition to these, as pointed out in \cite{Hertog:2004dr}, the asymptotic fall-off with $\beta \,=\, \tilde f \alpha^2$ for some $\tilde f$, may also be viewed as a one-parameter family of boundary conditions for a $\Delta=1$ operator ${\cal O}$ in the CFT with a triple trace deformation $\sim \tilde f {\cal O}^3$.
%Two of these quantisations can be identified as the ones which were originally discussed in \cite{kw}, wherein one of the two independent falloffs in $\Phi$ can be viewed as the mode dual to an operator of dimension $\Delta=1$ or $\Delta=2$, while the other ``normalisable'' mode yields the VEV of this operator. The one parameter family of allowed boundary conditions also allows the 
%asymptotic behaviour \eqref{asymptriple} the scalar to be interpreted as a deformation by a triple-trace marginal operator of the conformal fixed point where the bulk scalar is dual to a $\Delta=1$ CFT operator (Neumann boundary conditions) as in \cite{Hertog:2004dr}. 
We will, however, implicitly have in mind the natural interpretation suggested in \cite{maldacena1} that the boundary asymptotics corresponds to a relevant deformation by the $\Delta =1$ operator with a non-vanishing VEV for the same. The question of stability depends on the choice of interpretation/boundary conditions. In the paper, we will not focus on the issue of stability and interpretation. We choose this specific background as an analytically tractable setting within which to explore the properties of spacelike correlators of large dimension QFT operators that have geodesic limits in the bulk.
%Thus $f_0$ is the coefficient of the slowest fall-off and parametrizes the size of the deformation in the dual field theory.   

At the Euclidean origin $u\to\infty$, the corresponding asymptotics are
\bea
&&\Phi\,\simeq\,\Phi_0\,+\,{\cal O}(u^{-2})\,,\qquad\qquad
a(\xi(u))\,\simeq\,\frac{a_0}{u}\,+\,{\cal O}(u^{-3})\,,\\\nonumber\\\nonumber
&&\Phi_0\,=\, \sqrt{\tfrac{3}{2}}\, \ln{\left(\tfrac{f_0}{\sqrt{6}}+\sqrt{1\,+\,\tfrac{f_0^2}{6}}\right)}\,,\qquad
a_0\, =\,\sqrt{2}\left(1+\sqrt{1+\tfrac{f_0^2}{6}}\right)^{-1/2}\,.
\eea
Thus $f_0=0$ corresponds to undeformed AdS and the value of the field at the origin increases monotonically with $f_0$.   Note also that near the origin 
$u\to \infty$, the metric is smooth and flat, for all values of $f_0$.

\subsection{Lorentzian continuation} 
The analytic continuation to Lorentzian signature proceeds as usual by taking the polar angle on  $S^3$ to be the de-Sitter time coordinate $\theta = i t +\pi/2$.   The metric is
\begin{equation}\label{eqn:yiannisexteriormetric}
ds^2 \,=\, \left(1\,-\,\frac{f(u)^2}{6}\right)\,\left[\frac{du^2}{u^2(1+u^2)}\,+\,\frac{1}{u^2} (-dt^2\,+\,\cosh^2{t}\,d\Omega_2^2)\right]\,,
\end{equation}
which is asymptotically AdS$_4$ with global dS$_3$ slices.   This is valid in the exterior region $0 < u <\infty$.   The Euclidean origin $u\to\infty$ maps to a light cone through which we  continue into the interior FRW region by taking $u\,=\,-i v^{-1}$ and  $t\,=\,\chi-i\pi/2$ to get the metric,
\begin{equation}
ds^2\, =\, \left(1\,-\,\frac{f(-i v^{-1})^2}{6}\right)\,\left[-\frac{dv^2}{1-v^2}\,+\,v^2\, (d\chi^2+\sinh^2{\chi}\,d\Omega_2^2)\right]\,. 
\end{equation}
The horizon is at $v=0$ and increasing $v$ takes us  away from the horizon into the FRW region, while reality of the metric requires $v \leq 1$.  The Lorentzian continuation of the metric in these coordinates presents a subtlety: the  cosmological time $\sigma$, as a function of $v$, is multivalued. This is most clear when the deformation $f_0$ is turned off. In this situation, to recover the standard form of the AdS metric in the FRW patch, the cosmological time is given as $\sigma \,=\,\sin^{-1}v$ with $0\leq \sigma < \pi$. In this range, $v$ is positive (and less than unity), but $\sqrt{1-v^2}\,=\,\cos\sigma$ changes sign at $\sigma =\pi/2$ which corresponds to the branch point at $v=1$.

Upon turning on the deformation, the cosmological time $\sigma$ continues to be a multivalued function of $v$, defined via the integral
\be
\sigma\,=\, \int_0^v \left(1\,-\,\frac{f\left(-i v^{-1}\right)^2}{6}\right)^{1/2} \,(1-v^2)^{-1/2}\,\, dv\,.\label{defsigma}
\ee
Denoting the cosmological time at which $v(\sigma)$  attains its maximum value (unity) by $\sigma_s$,
\be
v(\sigma_s)\,=\,1\,,
\ee
when $\sigma > \sigma_s$, we are required to switch from one branch to the other of the square root in the denominator of $f(-i/v)$\,,
\be
f\left(-iv^{-1}\right)\, =\,\frac{f_0 }{\pm\sqrt{1-v^2}\,+\,\sqrt{1+f_0^2/6}}\,.
\ee
Close to the horizon, as $\sigma\to 0$, we must pick the positive sign which ensures that both the  FRW patch scale factor $\tilde a$ and the scalar field $\Phi$ are real, while for $\sigma > \sigma_s$ we must switch to the negative branch.

\begin{figure}
\centering
\includegraphics[width=2.5in]{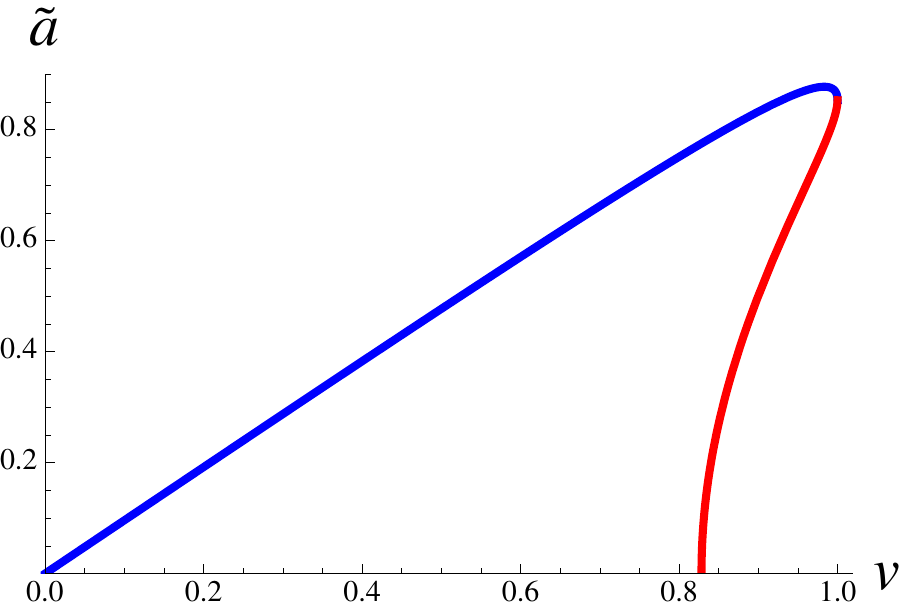}\hspace{0.3in}
\includegraphics[height=1.6in]{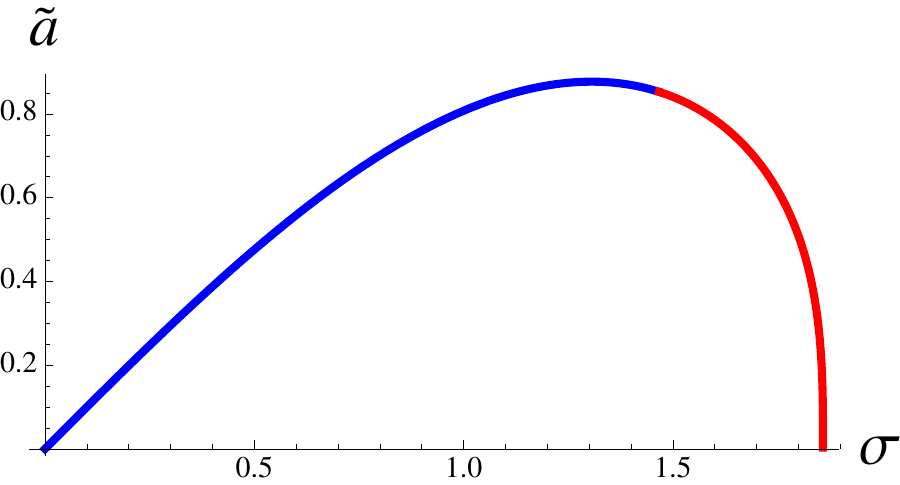}
\caption{\small{{\bf Left:} The scale factor $\tilde a$, in the FRW patch for the deformed AdS$_4$ solution with $f_0=1.5$ as a function of the $v$ coordinate. The scale factor is a double-valued function of $v$, with the two branches indicated in blue and red. {\bf Right:} The FRW scale factor as a function of cosmological time $\sigma$ with a maximum and a crunch.}}
\label{fig:ads4scale}
\end{figure}

Figure \ref{fig:ads4scale} shows the scale factor $\tilde a$ as a (multivalued) function of the coordinate $v$, and as single valued function of cosmological time $\sigma$. As expected on general grounds, the scale factor displays a maximum at $\sigma =\sigma_m < \pi/2$, and a crunch at  $\sigma = \sigma_c < \pi$.
The FRW scale factor $\tilde a$, 
\begin{equation}
\tilde{a} = \left(1-\frac{f(-iv^{-1})^2}{6}\right)^{1/2} v \,,\label{atilde}
\end{equation}
is vanishing at the horizon $v=0$ and increasing for small $v > 0$.  It is straightforward to find the location of the maximum and the crunch in terms of the $v$-coordinate. It turns out that the maximum is always located on the branch with positive sign for the square root in $f(-i/v)$ and occurs at $v=v_m$:
\bea\label{eqn:yiannismaximum}
&&v_m \,=\, \sqrt{\frac{2}{3}}\,f_0\, \left(C\,-\,\sqrt{2\,C}\right)^{-1/2}\,,\\\nonumber\\\nonumber
&&C\,\equiv\,1+\tfrac{1}{2}f_0^2+\sqrt{1+\tfrac{5}{3}f_0^2+\tfrac{1}{4}f_0^4}\,.
\eea
The location of the maximum\footnote{Note that the branch point at $v=1$ is not a maximum of the scale factor because $d\tilde{a}/d\sigma = (d\tilde{a}/dv)(dv/d\sigma)$ has a non-zero limit here.     
} satisfies the condition $v_m <1$, and exhibits the following asymptotic behaviours for small ($f_0\to 0$) and large ($f_0\to \infty$) deformations:
\bea
&&f_0\ll 1:\qquad\qquad v_m\,\simeq\,1\,-\,\frac{f_0^4}{72}\,,\qquad\qquad\qquad \tilde a_{\rm max}\,\simeq\,
1\,-\,\frac{f_0^2}{12}\,,\\\nonumber\\\nonumber
&&f_0\gg 1:\qquad\qquad v_m\,\simeq\,\sqrt{\frac{2}{3}}\,+\,\frac{1}{\sqrt{3} f_0}\,,\qquad
\qquad \tilde a_{\rm max}\,\simeq\,2^{5/4}\,\frac{1}{\sqrt{3 f_0}}\,.
\eea
The crunch singularity (the second zero of $\tilde a$) occurs when $f(-i/v)\,=\,\sqrt{6}$. At this point the scalar field  
$\Phi = \sqrt{6}\tanh^{-1}({f/\sqrt{6}})$ diverges. The location of the crunch is given by  $v=v_c$ with
\begin{equation}
v_c = \sqrt{2}\left(\tfrac{f_0}{\sqrt{6}}\right)^{1/2}\,\left(-\tfrac{f_0}{\sqrt{6}}\,+\,\sqrt{1+\tfrac{f_0^2}{6}}\right)^{1/2}\,.
\end{equation}
The crunch is located on the negative branch of the square root in $f(-i/v)$. 
It can be checked that curvature scalars of the metric diverge here. Importantly, using eq.\eqref{atilde} we infer that in the vicinity of the crunch
$\tilde a \sim \sqrt{(v-v_c)}$ and together with the definition of the cosmological time
\eqref{defsigma}, we obtain
\be
\tilde a(\sigma)\,\sim\, (\sigma -\sigma_c)^{1/3}\,,
\ee
in the near crunch regime. For the specific example discussed here, this result  validates the assumption made in section \ref{sec:outward} with regard to the exponent of the power law behaviour in the vicinity of the crunch singularity.

\subsection{Penrose diagram} 

We revisit the discussion in section \ref{sec:outward} in order to confirm, within the context of the deformed AdS$_4$ background, the validity of the general arguments presented there.  In particular, the shape of the crunch singularity (curving inwards or outwards) was shown to be determined by the real part of the tortoise coordinate at the crunch singularity.  The tortoise coordinate for our deformed AdS$_4$ background \eqref{eqn:yiannisexteriormetric} has a very simple form in terms of the radial variable $u$ in the exterior region,
\be
z \,\equiv\, \int_{\xi}^\infty \frac{d\xi}{a(\xi)} \,=\, -\int_u^0 \frac{du}{\sqrt{1+u^2}}\,=\, \sinh^{-1}{u}\,, \qquad\quad 0 <u < \infty\,,
\ee
since the conformal factor drops out.  As before, the boundary is at $z=0$, and the horizon at $z\to\infty$. Continuing past the horizon with $v\,=\, i u$ the tortoise coordinate in the FRW patch is then,
\bea
z\,=\,-\frac{i\pi}{2} \,+\, \ln{\frac{1\pm\sqrt{1-v^2}}{v}}\,,\qquad\qquad 0\leq v \leq 1\,.\label{zv}
\eea
Taking the positive branch of the square root, the horizon is approached from within the interior region as $z \to  -i\pi/2+\infty$. At $v=1$, the branch point of the square root corresponds to $z=z_s=-i\pi/2$, separating the location of the  maximum of the scale factor $z_m = z(v_m)$ from that of the crunch $z_c = z(v_c)$ with ${\rm Re}(z_c) < {\rm Re}(z_s) < {\rm Re}(z_m)$. 

Crucially, ${\rm Re}(z_c)$, which is evaluated with the negative sign for the square root in \eqref{zv}, is strictly negative. According to the discussion in section \ref{sec:outward}, this immediately implies that the spacelike crunch singularity is curved outwards as depicted in the Penrose diagram in figure \ref{fig:pda}.
 
\begin{figure}
\centering
\includegraphics[width=2.5in]{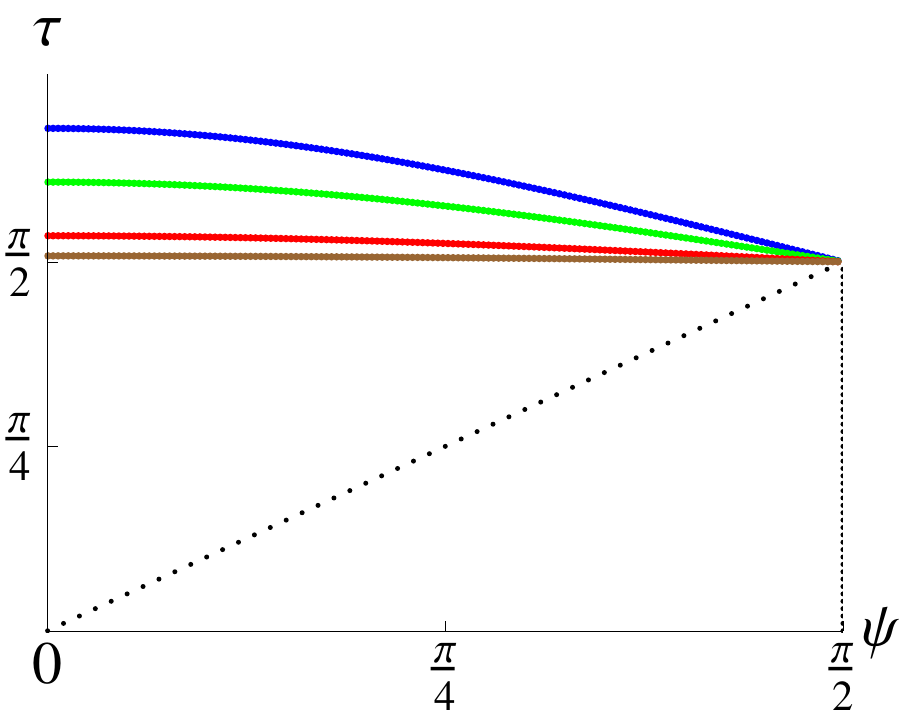}
\caption{\small{The outwardly curved spacelike crunch singularity plotted on the Penrose diagram for different values of the deformation, $g_0 =$ 0.3 (blue),
0.5 (green), 0.8 (red), 0.95 (brown). The conformal boundary is located at $\psi = \pi/2$.
As $g_0$ is increased towards $1$ (alternatively as $f_0\to\infty$), the crunch singularity approaches $\tau= \pi/2$. }}
\label{fig:crunches}
\end{figure}

Given the simple form of the deformed metric, we can proceed to obtain the Kruskal extension explicitly. For this purpose we suppress the spatial two-sphere in the geometry and focus attention on the two dimensional subspace spanned by the $t$ and $u$ coordinates. Using the substitution $u=\sinh{z}$ and subsequently the Kruskal null coordinates,
\begin{equation}
U\,=\, -e^{-t-z},\qquad V\,=\, e^{t-z}\,,
\end{equation}
the deformed AdS$_4$ metric \eqref{eqn:yiannisexteriormetric} becomes (suppressing angular coordinates),
\begin{equation}
ds^2\,
%\frac{a^2}{UV} 
=\, \left[-\frac{4}{(1+UV)^2} \,+\, \frac{4 g_0^2}{(1\,+\,g_0^2\, UV)^2}\right]\,dUdV\,. 
\end{equation}
The constant $g_0$ provides a convenient parametrization of the deformation, and is related to $f_0$ through
\begin{equation}
\frac{f_0}{\sqrt{6}}\, =\, \frac{2 g_0}{1\,-\,g_0^2}\,,\qquad\qquad 0 \leq g_0 < 1\,.
\end{equation}
Finally we compactify and straighten the boundaries using the coordinate change in eq.\eqref{straighten}
to yield 
\bea
&&ds^2= \Lambda^2(\tau,\psi)\sec^2\psi\,\left (-d\tau^2\,+\,d\psi^2\,+\,\sin^2\psi\,d\Omega_2^2\right)\,,\qquad\qquad 0\leq \psi < \frac{\pi}{2}\,,  \nonumber\\\\\nonumber
&&\Lambda^2 
\,=\, 1 \,-\, \frac{4g_0^2\,\cos^2\psi}{\left[(1+g_0^2)\,\cos{\psi} \,+\,(1-g_0^2)\,\cos{\tau}\right]^2} \,.
\eea
In these coordinates, the conformal boundary is at $\psi =\pi/2$ and corresponds to the cylinder (or Einstein static universe) ${\mathbb R}\times S^2$.  Since the crunch extends all the way to the boundary and intersects it at $\tau=\pi/2$, this is the ``end of time'' in the ESU picture.
The shape of the crunch singularity can be explicitly obtained  by plotting the zeroes of $\Lambda$ as displayed in figure \ref{fig:crunches}. The singularity is curved outwards as expected and flattens out towards $\tau =\frac{\pi}{2}$ as $g_0$ is dialled towards unity (equivalently, as $f_0\to \infty$).

\subsection{Radial geodesics}
As explained earlier, bulk global coordinates are regular (away from the crunch singularity) and convenient for describing spacelike geodesics whilst the dS-sliced coordinates are singular at the horizon. In global coordinates the conformal boundary is naturally the cylinder ${\mathbb R}\times S^2$ with an ``end of time'' at $\tau=\pi/2$. Unlike pure AdS, in the presence of the deformation, the bulk geometry is not static in global coordinates and there is no globally conserved energy. In dS-sliced coordinates, there is always a conserved de Sitter energy ${\cal E}$ which allows us to write down first order equations of motion (as in eq.\eqref{txifirstorder}). 

\begin{figure}
\centering
\includegraphics[width=2.8in]{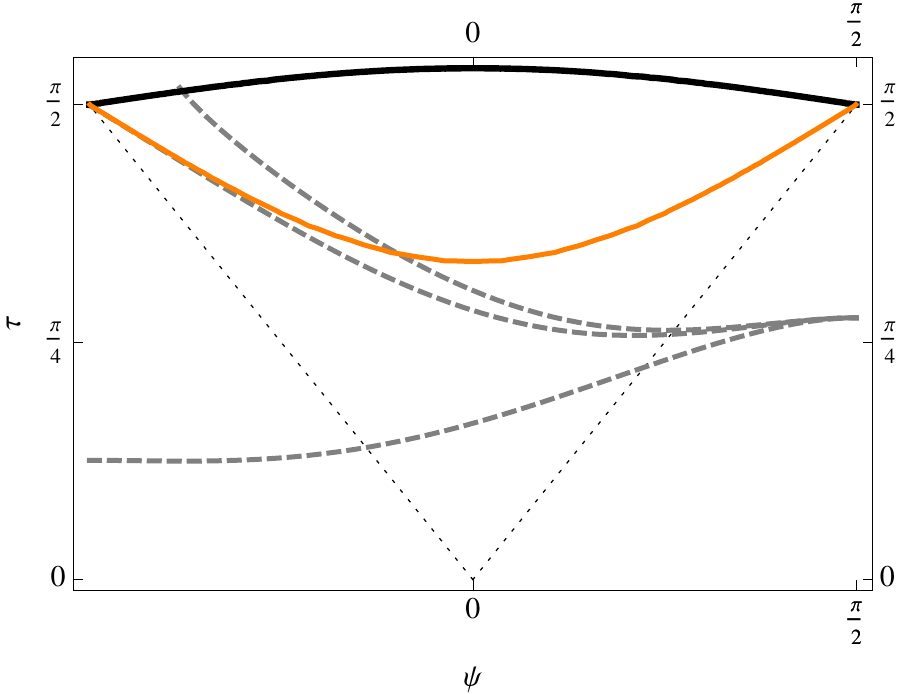}\hspace{0.3in}
\includegraphics[width=2.8in]{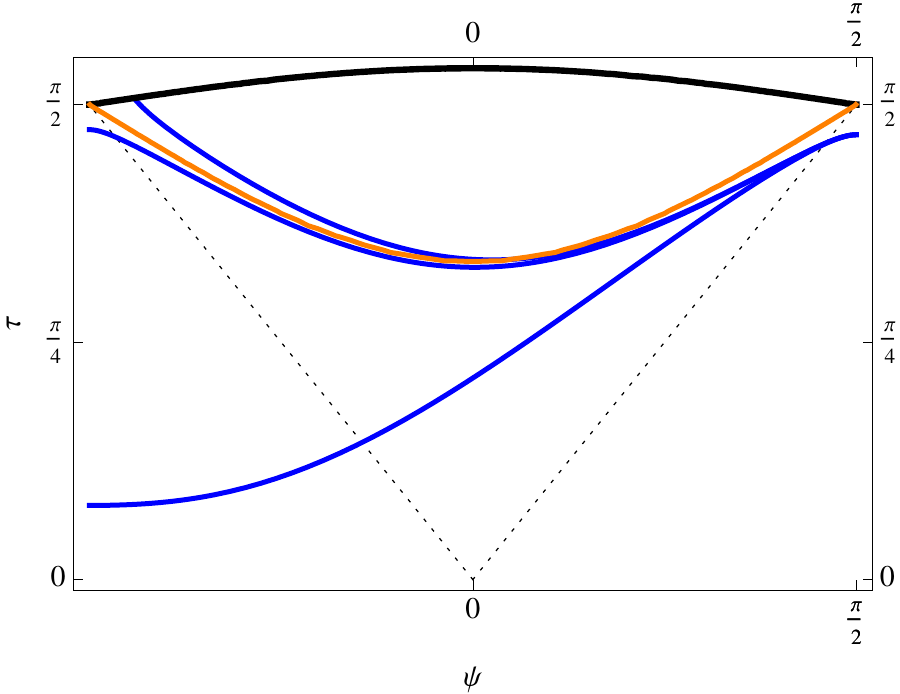}
\caption{\small{Spacelike geodesics in the deformed AdS$_4$ geometry with $f_0=10$ or $g_0\approx 0.785$. The big crunch is shown as a thick black line while the orange line marks the slice of maximal expansion behind the horizon $\tilde a \,=\, \tilde a_{\rm max} $. The dotted black lines represent the horizon. {\bf Left}: Three spacelike geodesics (dashed grey curves) are plotted with initial de Sitter time $t_1=0$ at de Sitter energies ${\cal E}\,=\,
0.7\,\tilde a_{\rm max},\,\tilde a_{\rm max}$ and $1.05\,\tilde a_{\rm max}$. {\bf Right:} Three spacelike geodesics (thick blue lines) for ${\cal E}\,=\,0.97 \,\tilde a_{\rm max},\, 0.9995\,\tilde a_{\rm max},\, 1.001\,\tilde a_{\rm max}$ and initial time $t_1=3$.}}
\label{fig:crunchinggeos}
\end{figure}

For the deformed geometry, the geodesic equations of motion do not appear to have closed form analytical solutions. Therefore, we first solve the equations of motion numerically in global coordinates for fixed de Sitter energy ${\cal E}$. This is achieved by making use of the fact that the geometry is asymptotically AdS$_4$ and that fixing $\tau(\psi)$ and $\tau'(\psi)$ at the conformal boundary (at one end-point of the geodesic) fixes the de Sitter energy. Specifically, we solve the second order equation of motion following from the action 
\be
S_{\rm ESU}\,=\, M\left(\int_0^{\pi/2 -\epsilon_1} \,+\int_0^{\pi/2 -\epsilon_2}
\right) 
\,d\psi \, \Lambda(\tau,\psi)\,{\rm sec}\,\psi\,\sqrt{1\,-\,\tau'(\psi)^2}\,\,,\label{actionesu}
\ee
subject to the boundary conditions ($\epsilon_{1,2}\ll 1$),
\be
\tau\,\left(\tfrac{\pi}{2}\,-\,\epsilon_{1,2}\,\right)\,=\,\tau_{1,2}\,\,,\qquad\qquad \tau'\left(\tfrac{\pi}{2}\,-\,\epsilon_{1,2}\,\right)\,\simeq\,-E_{1,2}\,\,\epsilon_{1,2}\,\,,\label{geodesicbc}
\ee
fixing the behaviour at each one of the two boundary endpoints which the geodesic is anchored to. The boundary behaviour is fixed by AdS asymptotics and the constant(s) of integration $E_{1,2}$ are related to the de Sitter energy ${\cal E}$ via
\be
{\cal E}\,=\,\sin\tau_{1,2}\,+\,E_{1,2}\,\cos\tau_{1,2}\,\,.\label{condition12}
\ee
The boundary time $\tau_1$ is related to the corresponding de-Sitter time by the boundary transformation $\tan{\tau_1} = \sinh{t_1}$. Integrating ``in'' from the first boundary point, fixing $\tau_1$ and varying ${\cal E}$ (or $E_1$, as dictated by eq.\eqref{condition12}), we numerically obtain a family of geodesics (see figure \ref{fig:crunchinggeos}). The global time and the asymptotic global energy $E_2$ at the second endpoint of each geodesic can be extracted from the solution using \eqref{geodesicbc} at the second endpoint
and it can be verified that the condition \eqref{condition12} is satisfied automatically.
A special exact solution is the constant de-Sitter time geodesic $\sin{\tau} = E/\sqrt{1+E^2}\sin{\psi}$ which has ${\cal E}=0$ and does not penetrate the bulk horizon and for which the asymptotic global energies on the two sides are equal $E_1=E_2=E$.  

Figure \ref{fig:crunchinggeos} clearly demonstrates that only the spacelike geodesics with de Sitter energies ${\cal E} < \tilde a_{\rm max}$  can go across the lightcone/horizon and connect the two antipodal points on the boundary. Solutions with ${\cal E} > \tilde a_{\rm max}$ have only one boundary endpoint and ``fall'' into the crunch singularity. Moreover, at late times $t_{1,2}\to \infty$ or $\tau_{1,2}\to \pi/2$, the geodesics hug the slice of maximal expansion with ${\cal E}={\tilde a}_{\rm max}$.

\subsubsection{Late time solution}
The fact that late time geodesics approach the maximal expansion slice in the FRW patch is crucial, as it allows us to obtain an analytical result for the corresponding correlator at large times. In order to calculate the geodesic action on this slice, we move back to the tortoise coordinate behind the horizon:
\be
z\,=\,-\frac{i\pi}{2}\,-\,w\,,\qquad\qquad \cos\tau\,=\,\tanh w\, \cos\psi\,,\qquad w\in {\mathbb R}\,.
\ee
The maximal expansion slice is a slice of constant $w$ satisfying
\be
\frac{\partial\tilde a}{\partial w}\,=\,\frac{\partial}{\partial w}
\left(
\frac{\Lambda(w)}{\cosh w}\right)\,=\,0\,.
\ee
This is also a constant-$w$ solution to the geodesic equations of motion, as can be straightforwardly checked. Explicitly, the value of $w =w_m$ at the maximal expansion slice is determined by the largest real root of the algebraic equation
\be
e^{8w_m}\, -\,(1+g_0^2)\, e^{6w_m}\,-\,6 g_0^2\, e^{4w_m}\, -\,g_0^2 (1+g_0^2)\, e^{2w_m} + g_0^4 \,=\, 0\,. 
\ee
For arbitrarily small and large deformations, $g_0\ll 1$, and $g_0\to 1$ (or equivalently, $f_0\ll 1$ and $f_0\to\infty$ respectively), the asymptotic values of $w_m$, the real part of the tortoise coordinate at the time of maximal expansion are given by,
\be
w_m\left.\right|_{g_0 \ll 1}\,\approx 4 g_0^2\,,\qquad\qquad
w_m\left.\right|_{g_0 \to 1}\,=\,\frac{1}{2}\ln\left(2+\sqrt{3}\right)\,+\,{\cal O}(1-g_0)\,.
\ee
In the undeformed, pure AdS geometry $w_m=0$ which corresponds to the global time slice $\tau=\frac{\pi}{2}$. When the deformation is turned on, the maximal expansion slice dips below $\tau=\frac{\pi}{2}$ as displayed in figure \ref{fig:crunchinggeos}.

Asymptotically near $\psi=\pi/2$ where the geometry approaches AdS$_4$  we should expect divergences in the geodesic action to be regulated by exactly the same subtractions that were performed for pure AdS. To understand how this works carefully, we note that the late time geodesic solution naturally splits in two pieces. The first portion lies on the slice of maximal expansion, up to some late global time specified by 
\be
\psi_{1,2}\,=\,\frac{\pi}{2}-\delta_{1,2}\,,\qquad\qquad \cos\tau_{1,2}\,=\,\tanh w_m\,\cos\psi_{1,2} \,,\qquad\qquad \delta_{1,2}\ll 1\,,
\ee
where we have explicitly written the relation between the tortoise coordinate behind the horizon and the global variables $(\psi,\tau)$, on the maximal slice.
As depicted in figure \ref{fig:crunchinggeos}, the second portion of the  geodesic exits the lightcone and reaches the boundary with vanishing derivative 
\be
\tau\left(\tfrac{\pi}{2}-\epsilon_{1,2}\right)\,\simeq\, \tau_{1,2}\,,\qquad\qquad
\tau'\left(\tfrac{\pi}{2}-\epsilon_{1,2}\right)\,\simeq\, - E\,\epsilon_{1,2}\,\,.
\ee
This approximation becomes increasingly accurate in the limit that the two times $\tau_{1,2}$ approach $\pi/2$, the so-called ``end of time''.

The action for the first portion of the late time geodesic that rests on the maximal slice behind the horizon is easily evaluated:
\be
S_{\rm ESU}^{(1)}\,=\, M\,\Lambda(w_m)\,{\rm sech}\, w_m\left(\int_0^{\frac{\pi}{2}-\delta_1}+\int_0^{\frac{\pi}{2}-\delta_2}\right)d\psi\,
\frac{\sec\psi}{\sqrt{1-\tanh^2 w_m \cos^2\psi}}\,.
\ee
The integration can be performed exactly and expressed in terms of the maximum value of the FRW scale factor as 
\be
S_{\rm ESU}^{(2)}\,=\, M\,\tilde a_{\rm max}\left[2\ln2 -\ln\delta_1-\ln\delta_2+2\ln\,\cosh w_m\right]\,.
\ee
The action for the second (infinitesimal) portion of the geodesic, that resides outside the horizon for late times, can be determined by setting $\tau=\tau_{1,2}$ and performing the integral \eqref{actionesu} from $\psi=\tfrac{\pi}{2}-\delta_{1,2}$ to $\psi=\tfrac{\pi}{2}-\epsilon_{1,2}$. We find
\be
S^{(2)}_{\rm ESU}\,=\,M\left(\ln \delta_1 + \ln \delta_2 - \ln\epsilon_1-\ln\epsilon_2\right)\,.
\ee
The divergent contributions proportional to $\ln \epsilon_{1,2}$ are precisely the same as in pure AdS spacetime and are removed by the same subtraction that is performed in AdS. On the cylinder ${\mathbb R} \times S^2$, the regulated action for the late time geodesic is,
\be
S_{\rm ESU,\,reg}\,=\,M\left[-(\tilde a_{\rm max}-1)\ln \delta_1\delta_2
\,+\,\tilde a_{\rm max}\ln 2\cosh w_m\right]\,.\label{latetreg}
\ee
Identifying $M$ with the conformal dimension of the dual operator $\Delta$ for large $\Delta$, we immediately obtain the late time correlator on the cylinder,
\bea
\langle{\cal O}_\Delta(\tau_1, \phi)\,{\cal O}_\Delta(\tau_2,\pi-\phi)\rangle_{\rm ESU}\,=&& e^{-S_{\rm ESU,\,reg}}\\\nonumber
&&\sim\,\left(\tfrac{\pi}{2}-\tau_1\right)^{\Delta(\tilde a_{\rm max}-1)}\,\left(\tfrac{\pi}{2}-\tau_2\right)^{\Delta(\tilde a_{\rm max}-1)}\,.
\eea
The corresponding result as a function de Sitter times $t_{1,2}$ can be derived by imposing the UV cutoff  at fixed radial coordinate $\xi =\xi_\infty$ in the dS-sliced  asymptotically AdS geometry (see eq.\eqref{dscutoff}):
\be
\langle{\cal O}_\Delta(t_1, \phi)\,{\cal O}_\Delta(t_2,\,\pi-\phi)\rangle_{\rm dS}\,\sim\,
e^{-\tilde a_{\rm max}(t_1+t_2)\Delta}\,.
\ee
Although we have focussed attention on the situation with both endpoints $t_{1}$ and $t_2$ large (or $\tau_1$ and $\tau_2$ approaching $\pi/2$), it is easily verified that the late time limit for any one of the two points leads to the same behaviour:
\bea
&&\left.\langle{\cal O}_\Delta(\tau_1, \phi)\,{\cal O}_\Delta(\tau_2,\pi-\phi)\rangle_{\rm ESU}\,\right|_{\tau_1\to \frac{\pi}{2}}\,\,\sim \,\,\left(\tfrac{\pi}{2}-\tau_1\right)^{\Delta(\tilde a_{\rm max}-1)}\,,\\\nonumber\\\nonumber
&&\left.\langle{\cal O}_\Delta(t_1,\phi)\,{\cal O}_\Delta(t_2,\,\pi-\phi)\rangle_{\rm dS}\,\right|_{t_1\to\infty}\,\,\sim\,\,
e^{-t_1\,\tilde a_{\rm max}\Delta}\,.
\eea
The late time asymptotics reveal certain key features. Boundary correlators on the cylinder exhibit a non-analyticity as $\tau_{1,2}\to \frac {\pi}{2}$. Since $\tilde a_{\rm max} < 1$, the non-analyticity is accompanied by a divergence in the correlator.   In contrast, for the undeformed CFT on ${\mathbb R}\times S^2$ (pure AdS), the limit $\tau_{1,2}\to \frac{\pi}{2}$ is smooth. 
 The only singularity in the CFT is a lightcone singularity (see \eqref{2ptesu}) when $\tau_1-\tau_2=\pi$. In the de Sitter space picture, the correlator decays exponentially at late times, with an exponent dependent on $\tilde a_{\rm max}$. In both pictures the correlator appears to factorise, suggesting that the late time behaviour is actually dictated by a one-point function for the operator ${\cal O}_{\Delta}$.

\subsection{Deformed AdS geodesics: general results}
The late time behaviour of the antipodal  correlator in the specific example above can be argued to be universal i.e. independent of  the details of the model. We have shown in section \ref{sec:generalities} that in a generic crunching AdS background, spacelike geodesics have a turnaround point behind the horizon
when ${\cal E}\,=\,\tilde a (\sigma)$. Therefore, geodesics with ${\cal E}> \tilde a_{\rm max}$ must end up at the crunch singularity behind the horizon, while the geodesic with ${\cal E}=\tilde a_{\rm max}$ lies entirely on the maximal expansion slice. In global bulk coordinates, the maximal expansion slice meets the boundary at the ``end'' of global time $\tau \to \frac{\pi}{2}$.  It is therefore natural that  late time geodesics $({\cal E} \lesssim \tilde a_{\rm max})$ in the dS-sliced picture which explore the FRW patch will remain arbitrarily close to the slice of maximal expansion. They only deviate from this slice close to the conformal boundary, when they must emerge into the exterior region to connect their boundary endpoints. Given the metric in the FRW patch,
\be
ds^2\,=\,-d\sigma^2\,+\,\tilde a^2(\sigma)\left(d\chi^2 + \sinh^2\chi \,d\Omega_{d-1}^2\right)\,,
\ee
the action for the geodesic on the maximal slice is 
\be
S\,=\,M\,\tilde a_{\rm max}\left(\int_0^{\chi_1} d\chi\,+\int_0^{\chi_2}d\chi\right)\,,
\ee
where $\chi_{1,2}$ are the UV cutoffs as the slice approaches the boundary on both sides (from within the FRW patch). Using the appropriate coordinate redefinitions (e.g. $\sec\psi\sin\tau=\sin\sigma\cosh\chi$), we can relate $\chi_{1,2}$ to the late time cutoffs:
\be
\chi_{1,2}\,=\,-\ln\left(\delta_{1,2}\right)\,-\,\ln\tanh w_m\,.
\ee
Defining the regulated action exactly as before, we reproduce the late time  nonalyticity ensuing from eq.\eqref{latetreg}.  It is also likely that the condition $\tilde a_{\rm max} < 1$ is a general feature. While this is true in the example we discussed in this paper, we have also found it to be the case in other, more complicated setups \cite{paper2}. For single scalar models it should be possible to argue this using the observation that $\ddot{\tilde a} < -\tilde a$ ( see footnote 5 for the origin of this inequality).

\section{Correlators and WKB limits}  
We have seen that correlation functions in the bulk geodesic limit do not directly probe the FRW crunch singularity behind the horizon.  In this sense the situation differs significantly from the AdS-Schwarzschild black hole \cite{shenkeretal}.
Although the late time behaviour of the correlators as a function of global time  shows a non-analyticity due to the ``end of time'' when the crunching surface intersects the conformal boundary, the strength of this particular non-analyticity  is only controlled by $\tilde a _{\rm max}$. The obvious question is whether  departures from the geodesic limit (or correlators of light operators) encode information on the crunch singularity in a subtle fashion. Since the solutions to the wave equations are not analytically tractable (with the exception of undeformed AdS), we need to understand how the WKB approximation can be implemented in the present context so that departures from the geodesic limit can eventually be explored. In this section, we take the first steps in this direction by making contact with the geometry behind the horizon in the WKB limit. The discussion closely follows the methods of \cite{fl1, fl2}.  In a separate discussion in appendix \ref{app:eads}, we  show how the calculation of the holographic  correlator in position space proceeds within the Euclidean  setting, and for the case of undeformed AdS, yielding the {\em exact} antipodal geodesic result after a WKB approximation followed by a steepest descent evaluation of frequency/momentum integrals.

\subsection{Scalar wave equation}
We consider a scalar field $\varphi$ of mass $m_\varphi$, dual to a boundary QFT operator ${\cal O}_{\Delta}$ with (UV) conformal dimension $\Delta$. The scalar $\varphi$ satisfies the free wave equation on the (dS-sliced) asymptotically AdS$_{d+1}$ background,
\be
ds^2\,=\, d\xi^2\,+\,a^2(\xi)\left(-dt^2 \,+\,\cosh^2 t\, d\Omega_{d-1}^2\right)\,.
\label{eqn:correlatorbackground}
\ee
To compute the QFT correlator, we need to solve the wave equation with the appropriate boundary conditions at the conformal boundary and at the origin (horizon). In the case of real time retarded correlators,  we would need to impose infalling boundary conditions  at the horizon\footnote{This approach was applied to the topological black hole which describes ${\cal N}=4$ SYM on dS$_3 \times S^1$ in \cite{Hutasoit:2009xy} to calculate retarded, real time correlation functions.} following the Son-Starinets prescription \cite{sonstarinets}.  However, we note that spatially antipodal points in global de Sitter spacetime are not causally connected \cite{Spradlin:2001pw}, and thus the non-vanishing antipodal correlators we have studied in the geodesic limit,  cannot correspond to the retarded boundary conditions. The issue is partially addressed if we define  correlation functions via analytic continuation from Euclidean  ones (see appendix \ref{app:eads} for the calculation in Euclidean AdS).
% We leave the implementation of precise boundary conditions that lead to different types of holographic correlators for future study. Below, we make some general observations on the frequency space Green's functions.

To solve the Klein-Gordon equation $(\Box -m^2)\varphi =0$ we use the manifest   symmetries (in Lorentzian signature) of the background to perform a separation of variables and express the scalar field as a mode expansion:
\begin{equation}
\varphi(\xi, t,\Omega)\, =\, \sum_{l m}\,Y_{l m}\left(\Omega\right)\,\int \frac{d\nu}{2\pi} \,\,\Xi_\nu(\xi)\,\,T_l(\nu,t)\,,\label{modeexp}
\end{equation}
where $Y_{lm}$ are spherical harmonics\footnote{The index $m$ stands for the collection of indices $\{m_i\}$ in the representation of $SO(d)$.  We use standard spherical harmonics.} on $S^{d-1}$, satisfying eqs.\eqref{ylmdef} and \eqref{ylmnormal}. The modes $T_l(\nu, t)$ solve a temporal mode equation \eqref{eqn:temporal} with eigenvalue $\sim \nu^2$, where $\nu$ plays the role of a frequency. They are given by solutions to the Schr\"odinger problem in a P\"oschl-Teller potential, and can be expressed in terms of associated Legendre functions as explained in detail in appendix \ref{app:waveeqn}. For even $d$, the temporal modes are simple, with ${\cal T}_l^\nu\,\sim\,P_\mu^{-i\nu}(\tanh t)$. Both the late time limit, and the high frequency limit of these modes yields plane waves, ${\cal T}_l^\nu\to \exp(-i\nu t)$. 

The deformation of the background affects the radial dependence through $a(\xi)$ and the radial wave equation  therefore plays the key role in  defining WKB-like limits. The radial part $\Xi_\nu$ of the  scalar field satisfies,
\begin{equation}
\Xi_\nu'' \,+\, d\,\frac{a'}{a}\,\Xi_\nu'\, +\, \left(\frac{\nu^2 +\tfrac{1}{4}(d-1)^2}{a^2}\, -\, m^2_\varphi\right)\,\Xi_\nu\, =\, 0\,,\label{eqn:radial}
\end{equation}
where the primes represent derivatives with respect to the radial coordinate $\xi$.
Recall that in pure AdS, the scale factor $a(\xi) = \sinh{\xi}$. This equation can be put in Schr\"odinger form by a rescaling of $\Xi_\nu$ and simultaneously switching to the tortoise coordinate $z$,
\begin{equation}
\Psi_\nu\, =\, a^{(d-1)/2}\,\Xi_\nu\,,\qquad\qquad z \,=\, \int_{\xi}^{\infty} \frac{d\zeta}{a(\zeta)}\,.
\end{equation}
Now the radial equation (\ref{eqn:radial}) is transformed into the Schr{\"o}dinger problem,
\bea
 -\frac{d^2 \Psi}{d z^2}\,+\, V(z) \Psi\,  =\, \nu^2 \,\Psi\,,
 \label{eqn:radialschrodinger}
\eea
with the potential function $V(z)$ defined as 
\be
V(z)\, =\, m^2_\varphi\, a^2\, +\, (d-1)\,\frac{1}{2a}\,\frac{d^2 a}{d z^2}\, +\,{(d-1)(d-3)}\,\frac{1}{4a^2}\,\left(\frac{da}{dz}\right)^2\,-\,\frac{1}{4}{(d-1)^2}\,.\label{eqn:schpotential}
\ee
For asymptotically AdS geometries, the conformal boundary is at $z=0$ where the Schr\"odinger potential should have a second order pole, $V\sim z^{-2}$. Near the horizon which is approached as $z\to \infty$, the function  $a(\xi(z))$ vanishes exponentially with $z$ and therefore the potential  $V(z)$ tends to a constant exponentially.
In particular, the undeformed AdS potential is,
\begin{equation}
V_{\rm AdS}\, =\,  \frac{(q^2-\frac{1}{4})}{\sinh^2{z}}\,,\qquad\qquad q\,\equiv\, \sqrt{\frac{d^2}{4}+m^2_\varphi}\,,
\end{equation}
so that the conformal dimension of the dual operator ${\cal O}_{\Delta}$ is $\Delta = d/2+q$.  
\subsection{WKB limit}
In order to make contact with the geodesic picture, we need to examine a WKB-like limit. 
This is the limit where the mass $m_\varphi$ of the bulk field is taken to be large, with frequencies and any non-vanishing angular momenta scaling in the same way with $m_\varphi$: 
\be
\nu\,=\, q\, u\,,\qquad \qquad \nu,\,q\to \infty\,,\quad u\,\,{\rm fixed}\,.
\ee
 Noting that the variable $q$ also scales as $m_\varphi$ in the limit of large mass, we write
\be
V(z)\,\equiv\,q^2\left(V_0(z)\,+\,\frac{V_2(z)}{q^2}\right)\,,
\qquad \qquad V_0(z)\,=\,a^2\,.
\ee
In the large $q$ limit, the Schr\"odinger potential is given by $V_0$  which depends only on $a(\xi(z))$.  Therefore the radial Schr\"odinger problem to be studied in the high frequency  WKB limit is 
\be
\left(-\frac{d^2 }{dz^2}\,+\, q^2\,a^2\right)\,\Psi_u^{\rm wkb}\,=\,q^2 u^2\,\Psi_u^{\rm wkb}\,.
\ee
The dependence of the Schr\"odinger potential on the scale factor $a$ has become remarkably simple in the WKB limit. The WKB potential $\sim a^2$ is monotonic (in the exterior region I), diverging as $z^{-2}$ for small $z$ and decaying exponentially $\sim e^{-2z}$ near the horizon $(z\to \infty)$.  
Therefore, for real frequencies $\nu$ (so that $u\in {\mathbb R}$), there is only one turning point at some $z=z_*$ such that
\be
u\,=\,a(z_*)\,,\qquad u \in {\mathbb R}\,.
\ee
The turning point separates the classically forbidden region $z<z_*(u)$ from the classically allowed region $z>z_*(u)$. We write the solutions in these two regions as, 
\bea
&&\Psi^{\rm wkb}_u(z)\,=\, \kappa^{-1/2} \left(A^-\, e^{- q \,\mathcal{Z}_u }\,+\,A^+\, e^{q \,\mathcal{Z}_u }\right)\,,\qquad z< z_*(u) \label{forbidden}\\\nonumber\\\nonumber
&&\mathcal{Z}_u(z) \,=\, \int_{z_*(u)}^{z} \kappa_u(z')\, dz'\,,\qquad\qquad
\kappa_u(z)\, =\, \sqrt{V_0(z) - u^2} \,,
\eea
while in the classically allowed region, we make the following replacements using standard WKB connection formulae as we go through the turning point:
\bea
&&\kappa^{-1/2}_u \,e^{q\, \mathcal{Z}_u} \, \longrightarrow \,\frac{2}{
\sqrt {\tilde{\kappa}_u}}\, \cos{\left( q\, \tilde{\mathcal{Z}}_u- \frac{\pi}{4}\right)}\,, \label{allowed}\\\nonumber
&&\kappa^{-1/2}_u\, e^{-q\, \mathcal{Z}_u} \, \longrightarrow\,  -\frac{1}{\sqrt{\tilde{\kappa}_u}} \,\sin{\left( q\, \tilde{\mathcal{Z}}_u- \frac{\pi}{4}\right)}\,, \\\nonumber
&&\tilde{\mathcal{Z}}_u(z) \, =\, \int_{z_*(u)}^{z} \tilde{\kappa}_u(z')\, dz'\,,\qquad\qquad\tilde{\kappa}_u(z) \,=\, \sqrt{u^2-V_0(z)} \,.
\eea
In the leading order WKB approximation, the phase integral ${\cal Z}_u$ in the forbidden region obeys the equation
\be
\boxed{\frac{{\cal Z}_u^{\prime 2}}{a^2}\,+\,\frac{u^2}{a^2}\,=\,1\,.}
\ee
This is to be compared with the first order constraint equation for the spacelike geodesics \eqref{firstorder},
\be
\left(\frac{d\xi}{d\lambda}\right)^2 \,-\,\frac{{\cal E}^2}{a^2}\,=\,1\,,
\ee
with ${\cal E}$ being the conserved energy variable. We are thus led to make the following identifications:
\be
\frac{{\cal Z}_u^{\prime}}{a}\,=\,\frac{d\xi}{d\lambda}\,,\qquad\qquad u\,=\, i{\cal E}\,.
\ee
Therefore, as expected, the proper velocity $\dot\xi$ of the geodesic maps to the derivative of the WKB phase. Furthermore, we learn that the precise relation of the geodesic energy ${\cal E}$ to the de Sitter mode frequency (in the WKB limit $q\simeq m_\varphi \gg 1$) is,
\be
\boxed{{\cal E}\,=\,-i \frac{\nu}{m_\varphi}\,.}
\ee
Hence, spacelike geodesics with real ${\cal E}$ correspond to modes with imaginary de Sitter frequency, and the turnaround point of the geodesic with $\dot \xi =0$ in the FRW patch should be equated to a WKB turning point for {imaginary} $\nu$, i.e.  when $u^2<0$.  
In the WKB language then, taking $u^2 = -{\cal E}^2$, the relevant turning point is a solution to the equation
\be
V_0(z)\,=\, a^2(z)\,=\,-{\cal E}^2\,,\qquad\qquad {\cal E}\in {\mathbb R}\,.
\ee
Clearly there are no solutions  on the real $z$-axis, and the roots must lie in the complex $z$-plane. In fact, we already know that this equation has solutions upon analytically continuing into the FRW patch where $z\,=\,-\frac{i\pi}{2}+w$ and $a^2\,=\,-\tilde a^2$ with $\tilde a, w\in {\mathbb R}$. In particular, since $\tilde a$ has a maximum in the FRW patch, turning points on this slice only exist if ${\cal E}< {\tilde a}_{\rm max}$ (when $u^2<0$).  This is the condition for the existence of spacelike geodesics connecting antipodal points on the boundary.
It is also important to recognize that one of the  turning points behind the horizon (for imaginary $u$) is the continuation of the unique turning point in the exterior region (for real $u$) to (small) imaginary
values of $u$.

WKB turning points in the complex plane are linked to the geometry behind the horizon.  The importance of these was underlined in the works of Festuccia and Liu \cite{fl1,fl2}, specifically within the context of  AdS-Schwarzschild black 
holes dual to thermal states in large-$N$ CFTs. Such turning points play an important role in the structure  of boundary correlators. 

\subsection{Retarded correlator and WKB limits}

The spatially antipodal de Sitter correlators that we have discussed in the geodesic limit in the bulk should most naturally be associated to Feynman or Wightman functions in the strongly coupled de Sitter space 
field theory. Calculating these from the bulk gravity picture requires the identification of the correct boundary conditions at the horizon. We postpone a careful analysis of these for future work. We can, however, readily calculate the retarded Green's function by employing the Son-Starinets prescription \cite{sonstarinets} which requires incoming boundary conditions at the horizon for the bulk field $\varphi$. This means that the frequency  harmonics of $\varphi$ (defined through the mode expansion\eqref{modeexp})  must satisfy 
\be
\left.\Xi_\nu\right|_{z\to \infty}\,\sim\, e^{i\nu z}\,.
\ee
The near-boundary behaviour of $\Xi_\nu(z)$ is fixed as usual by the AdS asymptotics of the background,
\be
\left.\Xi_\nu\right|_{z\to 0}\, =\, C_1\, z^{\frac{d}{2}-q}\, \left[1+ \mathcal{O}(z^2)\right]\,\, +\,\, C_2\, z^{\frac{d}{2}+q}\, \left[1+ \mathcal{O}(z^2)\right]\,,
\ee
where $C_1, C_2$ are integration constants constrained by the requirement that $\Xi_\nu=1$ at a UV regulating cutoff surface at $z=\epsilon$ so that 
\begin{equation}\label{eqn:radialsol}
\Xi_\nu = \frac{\Xi_{1,\,\nu}\, +\, \beta_{\nu}\, \Xi_{2,\,\nu}}{\epsilon^{\frac{d}{2}-q}\,+\,\beta_{\nu}\,\epsilon^{\frac{d}{2}+q}}\,,\qquad\qquad \beta_\nu\,\equiv\,\frac{C_2}{C_1}\,.
\end{equation}
$\Xi_{(1,2),\,\nu}$ represent the two linearly independent solutions with the two distinct types of AdS boundary asymptotics.
All non-trivial information on the (retarded) boundary correlator resides in the ratio $\beta_\nu$ viewed as a complex function of $\nu$. The solution to the WKB Schr\"odinger problem is related to the Fourier harmonic $\Xi_\nu$ according to
\be
\Psi_u^{\rm wkb}(z)\,=\,a^{(d-1)/2}\,\Xi_\nu\,.
\ee
Carefully using the WKB solutions \eqref{forbidden}, \eqref{allowed} in the classically forbidden and allowed regions, we find
\be
\beta_\nu\,=\,\frac{i}{2}\,\lim_{z\to 0}\,
\exp\left[2q\left(\int_{z_*(u)}^z \kappa_u(z')dz'\,-\,\ln z\right)\right]\,.
\label{betanu}
\ee
Finally, the frequency space retarded correlator is obtained by substituting the WKB solution for $\Xi_\nu$ into the boundary action and then using the orthonormality of spherical and temporal harmonic mode functions,
\be
S_{\rm bdry}\,\sim\,\int_{{\rm dS}_{d}} d^d x\, \sqrt{-g}\, g^{zz}\, \varphi\, \partial_{z}\left.\varphi \right|_{z=\epsilon} \,.
\ee
\begin{figure}
\centering
\includegraphics[width=2.7in]{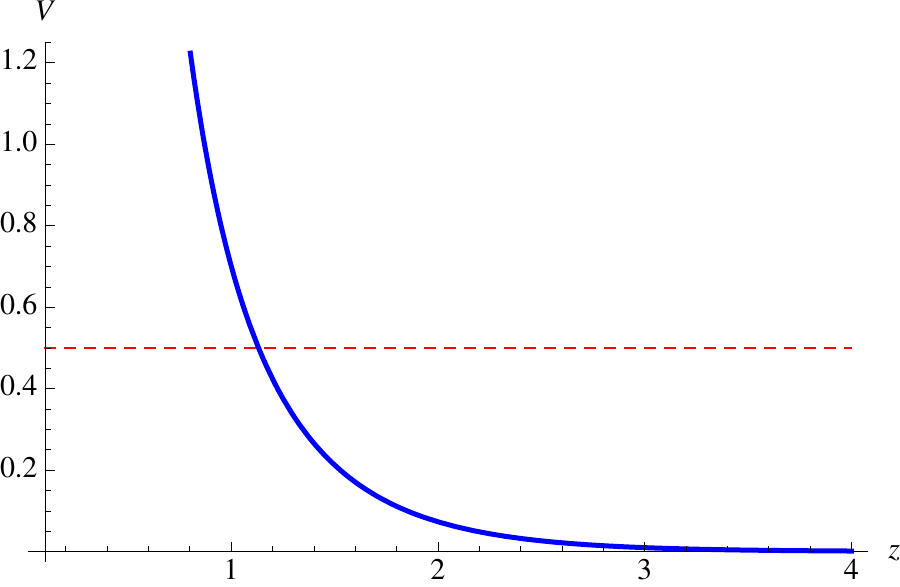}\hspace{0.2in}
\includegraphics[width=2.7in]{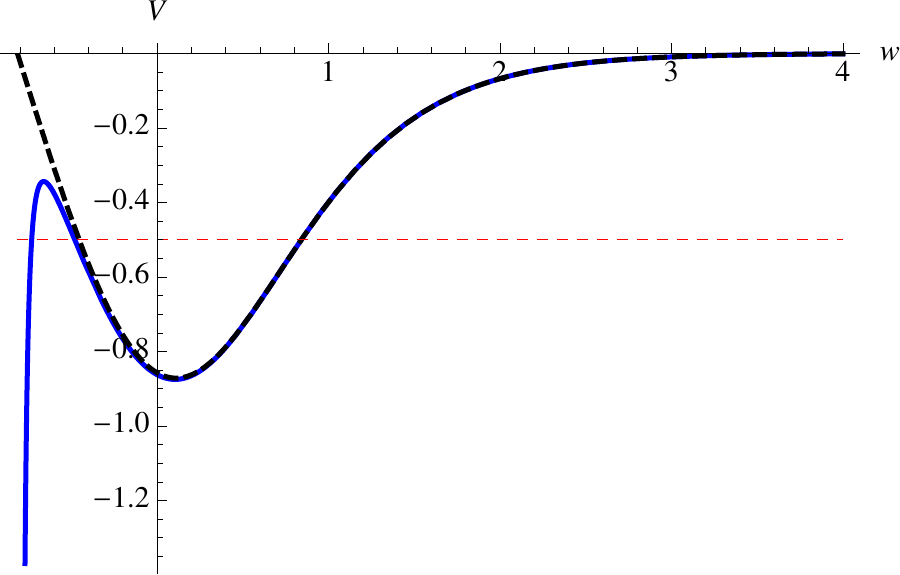}
\caption{\small{{\bf Left:} The Schr\"odinger potential $V(z)$ in the exterior region for the deformed AdS$_4$ geometry of section \ref{sec:yiannis} versus the tortoise coordinate $z$, for $f_0=1$. The dashed red line corresponds to a real frequency $\nu^2 =0.5$. {\bf Right:} The Schr\"odinger potential $V(z)$ in blue, plotted along the $w$-axis, the tortoise coordinate behind the horizon, $z=-\frac{i\pi}{2}+w$. The WKB potential $V_0$ behind the horizon is shown as the dashed black curve. The dashed red line corresponds to an imaginary frequency with $\nu^2=-0.5$. The qualitative difference between $V$ and $V_0$ is pronounced near the crunch, where $V\, \sim\, - (w -w_c)^{-2}$ whilst $V_0\, \sim\, (w-w_c) $.}}
\label{fig:wkbpotential}
\end{figure}
 %To compute the boundary term in the on-shell action we need 
%\begin{equation}
%\sinh^{-(d-1)}{(\epsilon)} \partial_z \Phi^{wkb}(\epsilon)  = i q \epsilon^{2q-%d} A^2
%\end{equation}
%up to contact terms.    We have the same calculation as before with $\frac{i}{2}A^2$ in place of $\beta_\nu$.   The temporal modes are just plane waves and the frequency space retarded Green's function is simply 
Up to additive contact terms, the frequency space Green's function in the WKB limit is given by 
\begin{equation}
\mathcal{G}^{\rm wkb}(u)\,=\,4  q \,\epsilon^{2q-d}\,\beta_\nu\,,
\end{equation}
where $\beta_\nu$ is defined as in \eqref{betanu}. For a generic deformed background, its exact form and singularities will depend on the scale factor $a$. However, we can draw some general inferences about the frequency space Green's function. Nontrivial dependence on the frequency $u = \nu/m_\varphi$ enters in \eqref{betanu} through the dependence of the turning point $z_*$ on $u$. While $z_*(u)$ is single-valued along the real $u$-axis, this is no longer the case when $u$ is complex. For strictly imaginary values of $u$ (so $u^2 <0$), there are two turning points $z_*$ satisfying
\be
a(z_*)^2\,=\, -{\cal E}^2\,, \qquad u\,=\, i {\cal E}\,.\label{turningpts}
\ee
This is because in the FRW patch the scale factor $\tilde a\,=\, -i a$ has a single maximum and therefore  eq.\eqref{turningpts} has two solutions when ${\cal E} < \tilde a_{\rm max}$. Of the two turning points, both of which lie on the line $z\,=\,-i\pi/2 + w\,,\, (w\in {\mathbb R})$, the one which is closer to the horizon is the ``physical'' turning point according to the terminology adopted in \cite{fl2} i.e. it is the analytic continuation of the unique WKB turning point in the exterior region for real $u$. Therefore $z_*(u)$ is not single-valued for imaginary $u$, and the two turning points merge precisely when ${\cal E}=\tilde a_{\rm max}$ signalling a branch point singularity.  

The complete analytic structure of ${\cal G}^{\rm wkb}(u)$ will clearly be determined by  the actual functional dependence of the WKB potential  $\sim a^2$ on $z$, the tortoise coordinate. This not only affects the WKB phase integral ${\cal Z}_u$, but also determines whether there are {\em other} branch points in the complex $u$-plane resulting from mergers of the physical turning point with other complex roots of $a(z_*)^2=u^2$. Regardless, the presence of two WKB turning points along the imaginary $u$-axis is always guaranteed by the fact that the FRW scale factor has a single maximum and two zeroes (at the horizon and the crunch singularity).

Therefore the retarded Green's function, which should have singularities only in the lower half plane, has a branch point singularity when
\be
\frac{\nu}{m_{\varphi}}\,=\,-i\,\tilde a_{\rm max}\,,\label{branchpt}
\ee
in the large mass or WKB limit.  The branch cut emanating from this branch point should extend away from the real axis (for retarded correlators). For the case of the AdS black hole, it could be argued in \cite{fl1,fl2} that  branch cuts such as this actually arise due to the infinite set of discrete quasinormal poles merging into a continuum in the WKB limit. By analogy then, we are led to identify the branch point \eqref{branchpt} with the lowest quasinormal mode (corresponding to a large dimension boundary operator). We can provide some supporting evidence for this expectation below by examining the same correlators for the undeformed theory i.e. the pure AdS geometry.

As already alluded to above, the correlator may well have other branch point singularities in the complex plane depending on the precise form of the scale factor $a$ and the number of roots for the WKB turning point condition $a(z_*)^2\,=\,u^2$ in the complex plane, for generic complex $u$. According to the criteria of \cite{fl2}, such singularities can be related to quasinormal modes if the branch points result from the merger of the analytic continuation of the physical turning point (defined above) with some other complex turning point. The exploration of the more general set of WKB turning points and their physical significance is a potentially interesting exercise that can be carried out for the analytically solvable AdS$_4$ deformation we have discussed in this paper.

\begin{figure}
\centering
\includegraphics[height=1.5in]{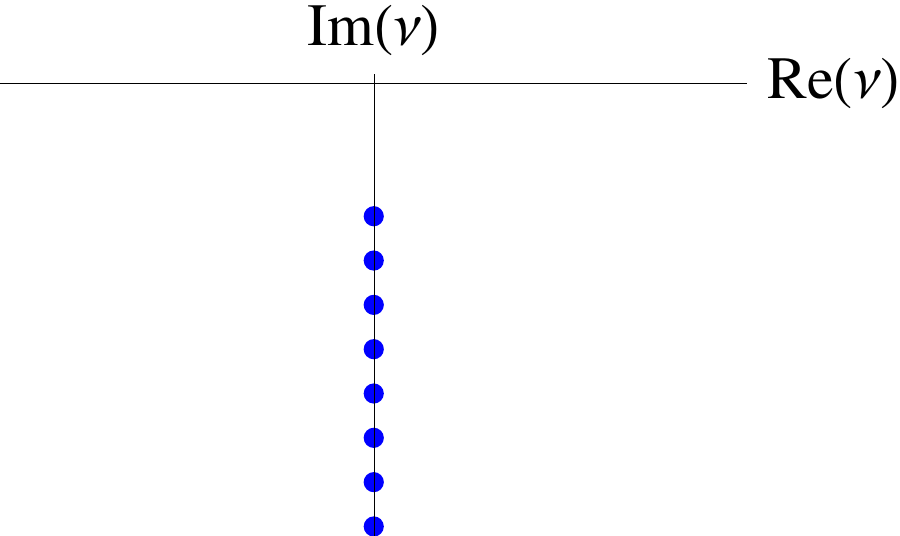}\hspace{0.5in}
\includegraphics[height=1.5in]{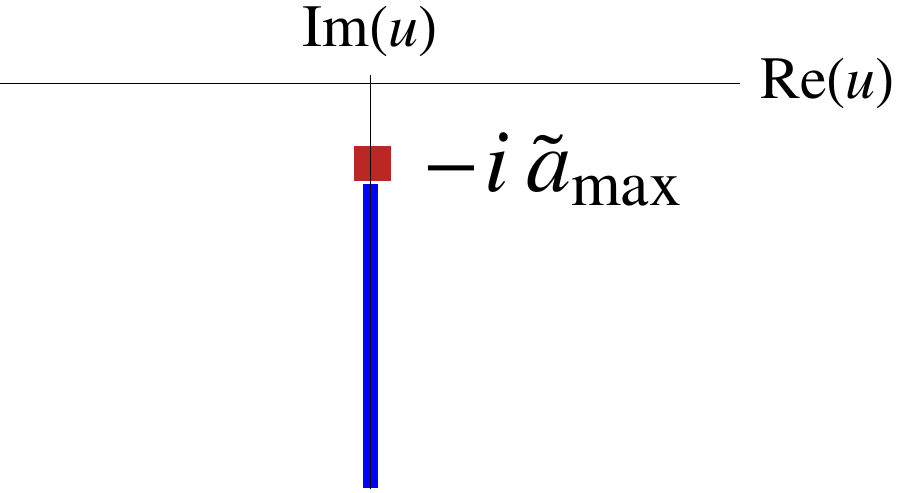}
\caption{\small{Quasinormal poles in the complex $\nu$-plane merge into a continuum on the $u$-plane $(u=\nu/q)$ in the large $q$ limit. The resulting branch point appears at $u \simeq -i \tilde a_{\rm max}$.}}
\label{qnm}
\end{figure}

\subsubsection{Pure AdS: a check}
\paragraph{The WKB result:} It can be verified that for pure AdS, the WKB approximation and the exact result for the retarded correlator are mutually consistent, and lend support to the  general picture described above. Using $a=1/\sinh z$ for the AdS geometry, we perform the WKB integral in eq.\eqref{betanu} exactly to obtain,
\be
{\cal G}^{\rm wkb}(u)\,\sim\,\exp\left[q \left\{ (iu +1)\ln(iu+1)-(iu-1)\ln(iu-1)\right\}\right]\,,
\ee
where we have omitted overall normalisation factors. This expression displays a branch point singularity at $u=-i$ which is in accord with the fact that the maximal FRW patch scale factor satisfies $\tilde a_{\rm max}=1$ for AdS spacetime. Although there also appears to be a branch point at $u=+i$, it is easy to establish that in the $q\to\infty$ limit, the branch cut emanating from it has vanishing discontinuity. This situation was also encountered and explained in detail in \cite{Hutasoit:2009xy}. Let us now confirm that the WKB Green's function above is indeed the high frequency limit of the exact AdS Green's function.

\paragraph{The exact AdS result:} 
For a fixed frequency $\nu$ the solution to the Schr\"odinger problem  in the undeformed AdS geometry can be written in terms of two independent hypergeometric functions $\Xi_{1,2}$ \eqref{eqn:hypergeo}. From the near-horizon expansion of the hypergeometric functions, and upon implementing the Son-Starinets prescription we  obtain (appendix \ref{app:radial})
\begin{equation}
\beta_\nu\, =\, \frac{2^{-2 q} \Gamma{[-q]}\,\Gamma{[\frac{1}{2}+q -i\nu]}}{\Gamma{[q]}\,\Gamma{[\frac{1}{2}-q-i\nu]}}\,.
\end{equation}
This can be viewed as the frequency space retarded correlator in the boundary CFT on de Sitter space \footnote{There are certain  subtleties relating to a finite, discrete set of temporal harmonics arising from  bound states in the P\"oschl-Teller potential; these do not, however, affect the form of $\beta_\nu$ quoted above.}.
 As a function of $\nu$, it is analytic in the upper half-plane, and has poles in the lower half plane at
\begin{equation}
\nu\, =\, -i\left(\frac{1}{2}+q+n\right),\qquad n=0,1,2,\ldots
\end{equation}
This is the spectrum of quasinormal frequencies in pure AdS$_{d+1}$ with dS-slicings, which should coincide with the spectrum of zero modes of the Laplacian for the field of mass $m_\varphi$ on Euclidean AdS$_{d+1}$  \cite{Denef:2009kn}. Recall that $q=\sqrt{d^2/4 +m_\varphi^2}$. The high frequency WKB limit then follows from Stirling's approximation ($\ln \Gamma(z)\simeq z\ln z$) which yields
\be
\beta_\nu \,\sim\, \exp\left[q \left\{ (iu +1)\ln(iu+1)-(iu-1)\ln(iu-1)\right\}\right]\,,
\ee
matching the direct WKB calculation above.

\subsubsection{Beyond WKB}
\begin{figure}
\centering
\includegraphics[width=2.8in]{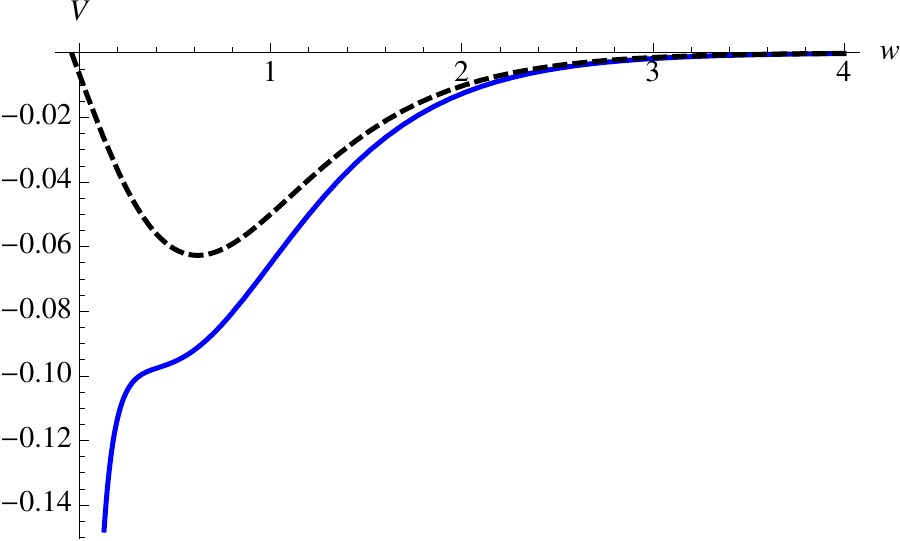}
\caption{\small{The Schr\"odinger potential (solid blue) and the WKB potential (dashed black) for the deformed AdS$_4$ model with deformation parameter $f_0=30$ and $q=10$. The WKB potential $V_0$ deviates significantly from the full potential $V$ which has only a single turning point when $u^2<0$. This happens whenever the deformation $f_0$ is dialled to large enough values.}}
\label{fig:singleturningpt}
\end{figure}

We have shown that the leading order WKB approximation probes the geometry behind the horizon through  complex turning points. However, the limit does not probe the big crunch directly. One immediate reason for this is evident from figure \ref{fig:wkbpotential} which shows the qualitative difference between the WKB potential $V_0\,=\,q^2\,a^2$ and the full Schr\"odinger potential $V(z)$ as a function of the tortoise coordinate in the FRW patch. For the single scalar truncation of ${\cal N}=8$ supergravity studied in section \ref{sec:yiannis} it turns out that the WKB potential $V_0$ vanishes linearly at the crunch while the formally subleading contribution $V_2$ causes the full Schr\"odinger potential to be unbounded from below at the crunch, where $V \sim - (w-w_c)^{-2}$. In fact it also turns out that for sufficiently large deformation parameter $f_0$, the local maximum in $V(z)$ completely disappears, so that $V(z)$ is monotonic and divergent at the crunch (see figure \ref{fig:singleturningpt}).

It is  possible to make a simple and general observation about the behaviour of the Schr\"odinger potential for a general scale factor that vanishes as $\tilde a(\sigma)\sim (\sigma-\sigma_c)^\gamma$ in the FRW patch\footnote{Recall that $\sigma$ is the continuation of the radial variable $\xi$ into the FRW region.} with $\gamma < 1$. Moving to tortoise coordinates, near the crunch we then have $a(w)=i\tilde a\sim i (w-w_c)^{\gamma/(1-\gamma)}$. Plugging into the complete Schr\"odinger potential we find that 
\be
\left.V\left(-\tfrac{i\pi}{2}+w\right)\right|_{w\to w_c}\,\sim\,\frac{(d-1)\left[\gamma(d+1)-2\right]\gamma}{4(1-\gamma)^2}\,(w-w_c)^{-2}\,.
\ee
The Schr\"odinger potential can thus diverge towards positive or negative infinity at the crunch depending on the value of $\gamma$ for a given $d$.
For the deformed AdS$_4$ model (where $d=3$ and $\gamma=1/3$), the potential diverges to negative infinity\footnote{Note that the inverse square potential also appears in the  asymptotically AdS exterior region. Obviously, this bears no relation to the inverse square behaviour in the vicinity of the crunch. Recall that for pure AdS where there is no crunch $w_c\to-\infty$.} as $-\frac{1}{4}(w-w_c)^{-2}$.
 This divergence and its effect on the next to leading order terms within the WKB approximation would be extremely interesting to understand. It is already interesting that the existence of the crunch singularity is reflected by an inverse square divergence in the Schr\"odinger potential for the radial wave equation.

Our discussion of frequency space correlators does not by itself establish a complete link to geodesics probing the FRW patch. In order to achieve this link we need to transform to a position space representation in the WKB limit. In addition, as a starting point, we need the correct prescription for Feynman or Wightman correlation functions. A somewhat different  route towards the same objective would be to first calculate Euclidean correlators between the antipodal points and subsequently analytically continue to  Lorentzian signature, thus yielding the desired boundary de Sitter space correlators. We initiate this study in appendix \ref{app:eads} and in  particular, we show how the correct position space correlator emerges from a saddle point method applied to a spherical harmonic spectral sum obtained by the standard holographic prescription in Euclidean AdS$_{d+1}$.

\section{Discussion}
We have tried to argue in this work that while geodesic probes of crunching AdS geometries do not get close to the singularity in the way that  analogous probes do for AdS black holes, there are interesting physical features encoded in their analysis. The most natural interpretation of the late time behaviour of the antipodal correlator in  Einstein static frame appears to be in terms of a divergent condensate for models with $\tilde a _{\rm max} <1$ which includes the specific AdS$_4$ model studied in this paper. It would be interesting to know if the upper bound on $\tilde a_{\max}$ is model independent; it appears to be true for the models we have looked at in this paper and in \cite{paper2}.
The unbounded growth of one-point functions or the CFT ``fall" has been argued \cite{Barbon:2011ta} to be a characteristic feature of the ``end of time'' in the Einstein frame in deformed CFTs with crunching duals; this includes deformations that have stable ground states in the dS-spacetime formulation.

It appears that the signature of the crunch singularity, if any, should be encoded in  frequency space Green's functions.  This is the immediate and most naive interpretation of the behaviour of the antipodal geodesics as a function of de Sitter energy ${\cal E}$. The fact that geodesics with energy ${\cal E} > \tilde a_{\rm max}$ terminate at the crunch, suggests a potential non-analyticity in frequency space correlators. Whether this non-analyticity can be identified in a meaningful way depends on a direct analysis of frequency space Green's functions of the (strongly coupled) field theory on the de Sitter background. For AdS black-holes the works of Festuccia and Liu \cite{fl1, fl2} showed how information on the singularity is encoded in the high frequency falloff of the Green's functions along the imaginary frequency axis. Within the context of this paper on the other hand, it appears  that if there is nontrivial information on the singularity, it must be encoded in the behaviour of the Green's functions for imaginary frequencies $|{\rm Im }\,\nu|/m_\varphi =\,|{\rm Im}\,u|> \tilde a_{\rm max}$. This is the indication given by the Schr\"odinger equation that the analytically continued wave equation reduces to for imaginary frequencies. The complex turning points of the WKB problem do imply a branch point at $|{\rm Im}\,u| = \tilde a_{\rm max}$. However, the behaviour of the Schr\"odinger potential for $|{\rm Im}\, u| > \tilde a_{\rm max}$ is not accurately captured within the WKB approximation; the WKB potential deviates significantly from the actual potential which is singular in the near-crunch region.  In fact, the full potential is not only singular but the number of its extrema in the FRW patch changes with $f_0$, the deformation parameter, as can be seen in figures \ref{fig:wkbpotential} and \ref{fig:singleturningpt}.  This is an indication  that next to leading order  effects in the WKB approximation will become essential. 

The Schr\"odinger potential for the deformed AdS$_4$ model appears to have several complex turning points for general complex $u$. It would clearly be of interest to learn if some of these turning points can merge with the ``physical'' turning point giving rise to new branch points (accompanied by their respective branch cuts) in the complex $u$-plane. 

 The calculation of holographic correlators is really a problem that deals with the wave equation outside the horizon i.e. the exterior region. Relating it to the Schr\"odinger-type potential behind the horizon is subtle and is most naturally understood within the context of complex WKB turning points and associated analytic continuations. It remains to be seen whether the interesting aspects of the analytically continued wave equation actually translate into concrete features in the frequency space Green's functions.

\acknowledgments 
We would like to thank Justin David and Ioannis Papadimitriou for useful discussions. We acknowledge financial support from the STFC grant awards  ST/L000369/1 and ST/K5023761/1.

%\startappendix
\newpage
\appendix
%\section*{Appendix}

\section{Transforming from dS-slicing to global coordinates}
\label{app:dstoglobal}
Recall that the Penrose diagram of  AdS spacetime ($\Lambda \,=\,1$) is  constructed starting from global coordinates ($r,\tau$)  which are related to the coordinates of the exterior (dS-slicing)  and interior (FRW)  patches as
\bea 
&&r\,=\, \sinh{\hat{\xi}}\, \cosh{t}\,=\, \sin{\hat{\sigma}}\sinh{\chi}\,, \label{eqn:patchglobal}\\\nonumber
&&\tan{\tau}\, =\,\tanh{\hat{\xi}}\, \sinh{t} = \tan{\hat{\sigma}}\, \cosh{\chi}\,.
\eea
The metric in these coordinates takes the globally static form
\begin{equation}
ds^2 \,=\,-(1+r^2)\,d\tau^2\, +\, (1+r^2)^{-1}\, dr^2\, +\, r^2\, d\Omega_{d-1}^2\,.
\end{equation}
The coordinate ranges are $r\geq 0$, $-\infty<\tau<\infty$,   i.e. we are taking the universal cover of AdS with $\tau$ unwrapped.    Further compactification $r = \tan{\psi}$ shows that AdS is conformal to the region $0 \leq \psi <\pi/2$ of the Einstein static universe ${\mathbb R}\times {\rm S}^d$
\begin{equation}\label{eqn:adsesu}
ds^2 = \sec^2{\psi} \left(-d\tau^2 +d\psi^2 +\sin^2{\psi}\, d\Omega_{d-1}^2 \right)\,.
\end{equation}

\section{Scalar wave modes}
\label{app:waveeqn}
The scalar field on the dS-sliced, asymptotically AdS geometry satisfies the Klein-Gordon equation, the general solutions to which are obtained by separation of variables:
\begin{equation}
\varphi(\xi, t,\Omega)\, =\, \sum_{l m}\,Y_{l m}\left(\Omega\right)\,\int \frac{d\lambda}{2\pi} \,\,\Xi(\lambda,\xi)\,\,T_l(\lambda,t)\,.
\end{equation}
The $Y_{lm}$ are the spherical harmonics on $S^{d-1}$ satisfying
\bea
\nabla^2_{S^{d-1}} \,Y_{l m} \, =\, -l(l+d-2)\, Y_{l m}\,,\label{ylmdef}
\eea
with the normalisation conditions
\bea
&&\int d\Omega\, Y^*_{lm}(\Omega) Y_{\bar{l}\bar{m}}(\Omega) \,=\, \delta_{l\bar{l}}\delta_{m\bar{m}} \label{ylmnormal}\\\nonumber 
&&\int d\Omega\,  \sum_{m}  Y^*_{l m}(\Omega) Y_{l m}(\Omega) \,=\, \frac{(2l+d-2)(l+d-3)!}{l! (d-2)! }\,.
\eea

\subsection{Temporal equation}
The temporal harmonics $T_l(\nu, t)$ satisfy the mode equation
\bea
\frac{1}{(\cosh t)^{d-1}}\partial_t\left[(\cosh t)^{d-1}\,\partial_t T_l(\nu,t)\right]\,+\,\frac{l(l+d-2)}{\cosh^2{t}}&&\,T_l(\nu,t)\,=
\label{eqn:temporal}\\\nonumber
&& -\,\left(\nu^2\,+\, \tfrac{1}{4}(d-1)^2\right)\,T_l(\nu,t)\,. 
\eea 
After a  rescaling, 
\begin{equation}
T_l\, =\, \left(\cosh t\right)^{-(d-1)/2}\,\, \mathcal{T}_l\,,
\end{equation}
the temporal equation is reduced to a Schr{\"o}dinger problem in a P{\"o}schl-Teller potential:
\be
 -\,\ddot{\mathcal{T}}_l\, -\, \frac{\mu(\mu+1)}{\cosh^2{t}}\,\mathcal{T}_l\, =\, \nu^2 \,\mathcal{T}_l\,, \qquad\qquad \mu \,=\, \frac{1}{2}(d-3)+l,\quad l=0,1,2 \ldots\label{eqn:poschlteller}
\ee
The solution to the Schr\"odinger problem above includes  a discrete set of negative ``energy" i.e. $\nu^2<0$  bound states,
\bea
&&-i\nu_n\, = \begin{cases}1,2,...,\mu &\mbox{integer $\mu$}  \\
\frac{1}{2},\frac{3}{2},...,\mu &\mbox{half-integer $\mu$} \end{cases} 
\\\nonumber\\
&&\mathcal{T}_l^n  \,=\, P_{\mu}^{i\nu_n}(\tanh{t})\,,
\eea
where $P_\mu^{i\nu}$ is the associated Legendre function.
The lowest energy state is proportional to $(\cosh t)^{-\mu}$ \footnote{The solutions can also be expressed in terms of Gegenbauer polynomials via the relation $C_{\mu-n}^{n+1/2}(x)\sim (1-x^2)^{-n/2}P_\mu^{-n}(x)$ (Abramowitz and Stegun \cite{abram} p.780).}.  The normalization of the discrete  modes is fixed by 
\begin{equation}
N_{l}^n \equiv \int \mathcal{T}_l^n(t)^2\, dt\, =\,- \frac{\Gamma{[1+\mu+i\nu_n]}}{i\nu_n\Gamma{[1+\mu-i\nu_n]}}\,.
\end{equation}
In addition to the finite set of discrete states, the P\"oschl-Teller potential possesses a continuum of positive energy ($\nu^2 >0$) scattering states\footnote{When $\mu=0$ these are simple exponentials $e^{\pm i \nu t}$, higher values of $\mu$ represent modulated exponentials e.g. $P^{i\nu}_{1}(\tanh{t}) \sim (1+ i \nu^{-1}\tanh{t})e^{i\nu t}$.}. The basis for these states  is provided by  two appropriate  linearly independent combinations of the associated Legendre functions $P_{\mu}^{i\nu}(\tanh{t})$, $Q_{\mu}^{i\nu}(\tanh{t})$, $P_{\mu}^{-i\nu}(\tanh{t})$, and $Q_{\mu}^{-i\nu}(\tanh{t})$.   The Legendre $P$ and $Q$ functions satisfy,
\begin{align}
P_{\mu}^{-i\nu}(\tanh{t}) &\,=\, \frac{\Gamma{[\mu-i\nu+1]}}{\Gamma{[\mu+i\nu+1]}}\left(\cosh{\pi \nu} \,P_{\mu}^{i\nu}(\tanh{t}) - \frac{2 i}{\pi}\sinh{\pi \nu} \,Q_{\mu}^{i\nu}(\tanh{t})\right) \nonumber \\\label{PQrelations}\\\nonumber
Q_{\mu}^{-i\nu}(\tanh{t}) &\,=\, \frac{\Gamma{[\mu-i\nu+1]}}{\Gamma{[\mu+i\nu+1]}}\left(\frac{i\pi}{2}\sinh{\pi \nu} \,P_{\mu}^{i\nu}(\tanh{t}) + \cosh{\pi \nu}\, Q_{\mu}^{i\nu}(\tanh{t})\right)\,.
\end{align}
The far future asymptotics ($t\to\infty$) of the modes are given as:
\begin{align}
P_{\mu}^{\pm i\nu}(\tanh{t})|_{t\to\infty} &= \frac{e^{\pm i\nu t}}{\Gamma{[1\mp i\nu]}} \nonumber \\
Q_{\mu}^{\pm i\nu}(\tanh{t})|_{t\to\infty} &= \frac{1}{2}\cosh{\pi\nu}\Gamma{[\pm i\nu]} e^{\pm i\nu t} + \frac{\sin{\pi(\mu\mp i\nu)}\Gamma{[-\mu\pm i\nu]} \Gamma{[\mp i\nu]} e^{\mp i\nu t}}{2\sin{\pi(\mu\pm i\nu)}\Gamma{[-\mu\mp i\nu]}}
\end{align}
We see that $P_{\mu}^{-i\nu}(\tanh{t})$ is always positive frequency in this limit.   We may choose to expand the temporal dependence in terms of these positive frequency modes in the ``far future basis'',
\begin{equation}
\mathcal{T}_l^{\nu}\, =\, \Gamma{[1+i\nu]} P_{\mu}^{-i\nu}(\tanh{t})\,,\label{devenbasis}
\end{equation}
where the normalisation is fixed so that in the limit $\nu\to\infty$ the modes reduce to the flat space ones:   
\begin{equation}
\lim_{\nu\to\infty}\mathcal{T}_l^\nu \,=\, e^{-i \nu t}  \quad \forall\, \mu\,.
\end{equation}
For $d$ odd, the $\{T_l^\nu\}$ can be shown to be orthonormal using the integrals presented in \cite{bielski}. When $d$ is even, however, an orthornormal basis is given by the set
\begin{equation}
\mathcal{\hat{T}}^\nu_l\, =\, \sqrt{\frac{\pi}{2}}\,\,\frac{\Gamma{[\frac{1}{2}+i\nu]} \,e^{\frac{1}{2}\pi |\nu|}}{\Gamma{[i\nu]}}  \left(P_{\mu}^{i\nu}(\tanh{t}) - \frac{2 i\, \sgn{(\nu)} }{\pi}\,Q_{\mu}^{i\nu}(\tanh\,t)\right)\,,\label{doddbasis} 
\end{equation}
\begin{equation}
\langle\mathcal{\hat{T}}^{\nu}_{l} \,\mathcal{\hat{T}}^{\nu'}_{l}\rangle\, \equiv\, \int_{-\infty}^{\infty}\mathcal{\hat{T}}^{\nu}_\mu (t)\,\mathcal{\hat{T}}^{\nu' }_\mu(t)\,dt\, =\, 2\pi\, \delta(\nu+\nu')\,.
\end{equation}
Modes with different values of $\mu$ are always orthogonal. 
\subsection{Radial solution in undeformed AdS}
\label{app:radial}
The radial wave equation \eqref{eqn:radial} for the scalar field $\varphi$ in the case of the pure AdS geometry is solved by two independent hypergeometric functions
\bea
&\Xi_1 \,=\, \left(\cosh z\right)^{\frac{1}{2}(d-1)+i\nu}  \left(\tanh z\right)^{\frac{d}{2}-q } \, _2F_1\left(\frac{1}{4} -\frac{1}{2}(q+i\nu),\frac{3}{4} -\frac{1}{2}(q+i\nu);1-q ;\tanh^2{z}\right)  \nonumber \\\label{eqn:hypergeo}\\\nonumber
&\Xi_2 \,=\, \left(\cosh z\right)^{\frac{1}{2}(d-1)+i\nu} \left(\tanh z\right)^{\frac{d}{2}+q } \, _2F_1\left(\frac{1}{4}+\frac{1}{2}(q-i\nu),\frac{3}{4}+\frac{1}{2}(q-i\nu);1+q ;\tanh^2{z}\right) \,,
\eea
which are well-defined for non-integer $q$\footnote{The integer case has additional poles which need to be handled separately.   There are also some other complications for half-integer $q$.  We will concentrate on the case of generic real $q>0$, as the final results should be expected to be insensitive to this issue.}. Near the horizon $\Xi_\nu = C_1 \Xi_1 + C_2 \Xi_2$ has the expansion,
\begin{align}
& \Xi_\nu(z\to\infty)\,=\,\frac{2^{-\frac{d}{2}-q}}{\pi^{3/2}}\, e^{\frac{1}{2}(d-1)z-i\nu z} \left( C_1 \cos{\pi(q+i\nu)}\,\Gamma{[1-q]}\,\Gamma{\left[\tfrac{1}{2}+q+i\nu\right]}\,\Gamma{[-i\nu]}\, + \right. \nonumber \\
&\left.+2^{2q} C_2 \cos{\pi(q-i\nu)}\,\Gamma{[1+q]}\,\Gamma{[\frac{1}{2}-q+i\nu]}\,\Gamma{[-i\nu]} \right) +(\nu\leftrightarrow -\nu) 
\end{align}
The radial functions diverge as $e^{\frac{1}{2}(d-1)z}$ but provided Im$(\nu)>-1/2$ they are still normalizable with the measure $\sqrt{-g} \sim a^d \sim e^{-d z}$.   In accord with the Son-Starinets prescription, we choose the ratio $C_1/C_2$ such that the solution goes as $e^{i\nu z}$ because this represents an incoming wave at the horizon.   

\section{Euclidean AdS correlators and WKB}
\label{app:eads}
Euclidean de Sitter spacetime is isomorphic to the $d$-sphere $S^d$. Specifying a point on $S^d$ by polar coordinates $(\theta,\,\Omega_{d-1})$, where $\Omega_{d-1}$ stands for coordinates on the $S^{d-1}$ latitude, the Euclidean correlator between two such points must have the form,
\begin{equation}
G_E(\theta_1,\theta_2,\delta\Omega_{d-1})\, =\, g(Z_{12})\,,
\end{equation}
where $Z_{12}$ is the $SO(d+1)$-invariant geodesic separation,
\begin{equation}
Z_{12} \, =\, \cos{\theta_1}\cos{\theta_2} \,+\, \sin{\theta_1}\sin{\theta_2}\cos{\delta\Omega_{d-1}}\,,\qquad \qquad Z_{12}\in [-1,1]\,. 
\end{equation}
To continue to $dS_d$ we take $\theta = i t+\pi/2$ yielding,
\begin{equation}
Z_{12} \,=\, -\sinh{t_1}\sinh{t_2} +\cosh{t_1}\cosh{t_2} \cos{\delta\Omega_{d-1}}\,, 
\end{equation}
which is now invariant under the de Sitter group $SO(d,1)$ with $Z_{12}\in\mathbb{R}$.   In particular when the two points are 
coincident or lightlike separated $Z_{12}=1$, and CFT correlators are typically rendered singular.   
We are primarily interested in the case where the two points are antipodal on the spatial sphere $S^{d-1}$.  In this instance $Z_{12} = -\cosh{(t_1+t_2)}$ and the two-point correlator of an operator with conformal dimension $\Delta$ is fixed by conformal invariance to be
\be
{\cal G}^{12}_{\Delta}\,\sim\,\frac{1}{(1-Z_{12})^\Delta}\,=\,
\frac{1}{\left[2\cosh^2\left(\frac{t_1+t_2}{2}\right)\right]^\Delta}\,.
\ee     

\subsection{Correlator on $S^d$ from holography}
Now we wish to understand the position space CFT result from a holographic standpoint. This requires solving the wave equation 
$\Box\phi-m^2\varphi=0$ on the Euclidean background
\begin{equation}
ds^2\, =\, a^2(z)(dz^2 +d\Omega_d^2)\,,
\end{equation}
where $z$ is the tortoise coordinate. For EAdS$_{d+1}$, the function $a(z)=1/\sinh z$. We then expand the field $\varphi$ in $S^d$ spherical harmonics,
\begin{equation}
\varphi \,=\, \sum_L \,Y_L(\Omega)\, \Xi_L (z)\,. 
\end{equation}
$Y_L$ is the spherical harmonic on $S^d$, satisfying
\begin{equation}\label{eqn:angular}
\nabla^2_{S^{d}} Y_L = -L(L+d-1) Y_L,\qquad L=0,1,2 \ldots
\end{equation}
The rescaled field $\Psi_L \,=\, a^{(d-1)/2}\Xi_L $ solves the radial equation 
\begin{align}
& -\frac{d^2 \Psi_L}{d z^2}+ V \Psi_L\,  =\, -\omega^2 \Psi_L\,, \nonumber \\
&V\, =\, m_\varphi^2 a^2 + \frac{d-1}{2 a}\frac{d^2 a}{d z^2} +\frac{(d-1)(d-3)}{4a^2}\left(\frac{da}{dz}\right)^2-\frac{(d-1)^2}{4}\,,
\end{align}
where $\omega = L+(d-1)/2$. In pure AdS the solution of the radial equation is
\begin{equation}
\Psi_L\, =\, C_1 P_{q-\frac{1}{2}}^{-\omega}(\coth{z}) + C_2 P_{q-\frac{1}{2}}^{\omega}(\coth{z})\,.
\end{equation}
In EAdS we need solutions that are regular at the origin $z\to\infty$. Only the first term is regular in this limit and scales as $e^{-\omega z}$.    It has the following expansion near the boundary
\begin{equation}
\Psi_L(z\to 0) = C_1 i^\omega \sqrt{2} \left( \frac{2^{-q} \Gamma{[2q]}}{\Gamma{[\frac{1}{2}+q]}\Gamma{[\frac{1}{2}+q+\omega]}}z^{\frac{1}{2}-q}+\frac{2^q \Gamma{[-2q]}}{\Gamma{[\frac{1}{2}-q]}\Gamma{[\frac{1}{2}-q+\omega]}}z^{\frac{1}{2}+q}\right)\,.
\end{equation}
To compute the momentum space correlator, the regular solution $(C_2=0)$ must equal $1$ at $z=\epsilon$ (near the conformal boundary). Normalizing appropriately, the near boundary expansion is then 
\begin{equation}
\Xi_L \,=\, \frac{z^{\frac{d}{2}-q}+\beta_L z^{\frac{d}{2}+q}}{\epsilon^{\frac{d}{2}-q}+\beta_L \epsilon^{\frac{d}{2}+q}}\,,
\end{equation}
where
\begin{equation}
\beta_L \,=\, \frac{2^{-2 q} \Gamma{[-q]}\Gamma{[\frac{1}{2}+q+\omega]}}{\Gamma{[q]}\Gamma{[\frac{1}{2}-q+\omega]}}\,.
\end{equation}
In the deformed case a similar treatment will apply, except that the $\beta_L$ will be different.   Therefore, in position space, the most general solution of the wave equation, correctly normalized  at the boundary and regular at the origin, is 
\begin{equation}
\varphi(\Omega,z)\, \,=\,\sum_{LM} A_{LM} Y_{LM}(\Omega) \Xi_L (z) \,,
\end{equation}
where we have introduced the Fourier coefficients $\{A_{LM}\}$.     Since $Y_{LM}^* = (-1)^M Y_{L,-M}$ we also require $A_{LM}^* = (-1)^M A_{L,-M}$.   The scalar action reduces on shell to the surface terms 
\begin{align}
S &= \int d\Omega\, \sinh^{-(d-1)}{(z)} \,\varphi \,\partial_z\varphi |_{\epsilon}^{\infty} \nonumber \\ 
& = \sum_{L,M,\bar{L},\bar{M}} A_{LM} A^*_{\bar{L}\bar{M}} \int d\Omega\, Y_{LM}(\Omega) Y^*_{\bar{L}\bar{M}}(\Omega) \sinh^{-(d-1)}{(z)} \,\Xi_L \,\partial_z\Xi_{\bar{L}} |_{\epsilon}^{\infty}\,.
\end{align}
There is no contribution from the origin provided $\omega>-(d-1)/2$ which always holds.   For the boundary contribution 
\begin{equation}
 \Xi_L (\epsilon)=1, \qquad \left(\sinh \epsilon\right)^{-(d-1)}\, \Xi'_{\bar{L}}(\epsilon) = 2 q \epsilon^{ 2 q-d} \,,\beta_{\bar{L}}   
\end{equation}
discarding contact terms.    Using orthogonality of the spherical harmonics we have
\begin{equation}
S = 2 q \epsilon^{2q-d} \sum_{LM} A_{LM} \beta_L A^*_{LM}\,. 
\end{equation}
To determine the position space correlator, we functionally differentiate the action twice with respect to the source 
\begin{equation}
\varphi_b(\Omega) =\sum_{LM} A_{LM} Y_{LM}(\Omega)\,.
\end{equation}
Inverting this relation, the functional derivative is $\delta A_{ L M} /\delta \varphi_b(\Omega)  = Y_{LM}^*(\Omega)$ and the Euclidean correlator in position space is,
\begin{align}
{\cal G}_E(\Omega,\Omega') &= 2 q\epsilon^{2 q-d} \sum_{LM} \beta_L Y^*_{LM}(\Omega) Y_{LM}(\Omega') \nonumber \\
&= 2 q\epsilon^{2q-d} \sum_L \beta_L W_L(\delta\Omega)\,.
\end{align}
Here, ee have made use of  the spherical harmonic sum rule to obtain $W_L$:
\begin{equation}
W_L\,= \sum_M Y^*_{LM}(\Omega) Y_{LM}(\Omega') = \frac{(2 L+ d-1)}{(d-1)\Omega_{d}} C_L^{\frac{d-1}{2}}(\cos{\delta\Omega})\,,
\end{equation}
with $C$ a Gegenbauer polynomial, $\delta\Omega$ the angle between the two directions $\Omega$ and $\Omega'$ and $\Omega_d = \int d\Omega$ the total solid angle.     The final expression for the holographic correlator on $S^d$ is
\begin{equation}
\boxed{{\cal G}_E(Z) = \frac{2q \epsilon^{2q-d}}{(d-1)\Omega_d} \sum_{L=0}^\infty \beta_L (2L+d-1) C_L^{\frac{d-1}{2}}(Z)\,,}
\end{equation}
where $Z=\cos{\delta\Omega}$ is the $SO(d+1)$-invariant separation between the two points. This is a completely general expression for the Euclidean correlator. All nontrivial information on the deformations is contained in the function $\beta_L$.  In the conformal case $\beta_L$ is given by the expression 
\begin{equation}
\beta_L = \frac{2^{-2 q} \Gamma{[-q]}\Gamma{[L+q+\frac{d}{2}]}}{\Gamma{[q]}\Gamma{[L-q+\frac{d}{2}]}}\,,
\end{equation}
and the sum is symmetric under the reflection $L \to -(L+d-1)$.   

\paragraph{Special case ($d=2$):}
When $d=2$ the Gegenbauer functions reduce to ordinary Legendre polynomials and we have 
\begin{equation}
{\cal G}_E(Z) = 2^{1-2q} q \epsilon^{2q-2} \sum_{L=0}^{\infty}\frac{\Gamma{[-q]}}{\Gamma{[q]}}
 \frac{\Gamma{[L+q+1]}}{\Gamma{[L-q+1]}} (2L+1) P_L(Z) \,.
\end{equation}
Using the known result
\begin{equation}
\left(\frac{2}{1-x}\right)^{q+1} = \sum_{L=0}^{\infty} \frac{(2L+1)\Gamma{[-q]}\Gamma{[L+q+1]}}{\Gamma{[1+q]}\Gamma{[L-q+1]}} P_L(x) \,,
\end{equation}
which is divergent for $q>-1/4$, we get
\begin{equation}
{\cal G}_E(Z) = \frac{2^{2-q} q^2 \epsilon^{2q-2}}{(1-Z)^{q+1}}\,,
\end{equation}
consistent with the result from conformal invariance. Recall that the conformal dimension $\Delta = q + d/2$.

\subsection{WKB limit of holographic correlator on $S^d$}  
Now we examine the WKB limit, by taking the angular momenta $L$ and mass $m_\varphi$ to scale to infinity in the same way:
\begin{equation}
\omega\,=\, q w\,,\qquad L = q w-\frac{d-1}{2},\qquad q\to\infty\,. 
\end{equation}
The discrete sum over $L$ becomes an integral over $w$ and   the (holographic) correlator on $S^d$ in the conformal case can be expressed as 
\begin{equation}
{\cal G}_E(Z) = \tilde{\mathcal{N}} \sum_{L=0}^\infty  \frac{\Gamma{[L+q+\frac{d}{2}]}(2L+d-1)}{\Gamma{[L-q+\frac{d}{2}]}} C_L^{\frac{d-1}{2}}(Z)\,, 
\end{equation}
where $\tilde{N}$ is a normalisation: 
\begin{equation}
\tilde{\mathcal{N}} = \frac{2^{1-2q} q \epsilon^{2q-d}}{(d-1)\Omega_d }\frac{\Gamma{[-q]}}{\Gamma{[q]}} \,.
\end{equation}
At fixed $Z$ and $q$ the summand diverges at large $L$ as $L^{2q+\frac{d-1}{2}}$.   Inserting a regulator $e^{-L\zeta}$ to make the sum converge we can write, 
\begin{equation}
{\cal G}_E(Z)\, =\, \lim_{\zeta\to 0}\, \tilde{\mathcal{N}}\sum_{w=\frac{d-1}{2q}}^{\infty} e^{-q w \zeta}\, \frac{\Gamma{[q w+q+\frac{1}{2}]}(2q w)}{\Gamma{[q w-q+\frac{1}{2}]}}\, C_{q w-\frac{d-1}{2}}^{\frac{d-1}{2}}(Z)\,, 
\end{equation}
where the increments in $w$ are $\delta w = 1/q$. Using Stirling's approximation,
\begin{align}
\frac{\Gamma{[q w+q+\frac{1}{2}]}}{\Gamma{[q w-q+\frac{1}{2}]}}  & \sim  e^{ q X(w)}\,, \nonumber \\
X(w) &= (w+1)\ln{(w+1)}-(w-1)\ln{(w-1)}\,,
\end{align}
and letting $Z=\cos{\phi}$, the correlator can be represented as the integral:
\begin{equation}
{\cal G}_E^{\rm wkb}(\cos{\phi})\, \sim\, \int_0^{\infty} dw\, \cdot w\,\cdot  e^{q X(w)}\,\cdot C_{qw-\frac{d-1}{2}}^{\frac{d-1}{2}}(\cos{\phi})\,.
\end{equation}
Next we make use an integral representation of the Gegenbauer function:
\begin{equation}
C_L^{\lambda}(\cos{\phi}) = \frac{\Gamma{[L+2\lambda]}}{2^{2\lambda-1}\Gamma{[L+1]}\Gamma^2{[\lambda]}}\int_0^\pi d\theta\, (\cos{\phi}+i\cos{\theta}\sin{\phi})^L \sin^{2\lambda-1}{\theta}\,,
\end{equation}
which yields
\begin{align}
C_{qw-\frac{d-1}{2}}^{\frac{d-1}{2}}(\cos{\phi}) & =\frac{\Gamma{[q w+(d-1)/2]}}{2^{d-2}\Gamma{[qw-(d-3)/2]}\Gamma^2{[(d-1)/2]}} \int_0^\pi d\theta\, e^{-(q w-\frac{d-1}{2}) f(\theta)} \sin^{d-2}{\theta} \nonumber \\
& \approx \frac{(q w)^{d-2}}{2^{d-2}\Gamma^2{[(d-1)/2]}} \int_0^\pi d\theta\, e^{-(q w-\frac{d-1}{2}) f(\theta)} \sin^{d-2}{\theta}\,,
\end{align}
with
\begin{equation}
f(\theta) = -\ln{(\cos{\phi}+i\cos{\theta}\sin{\phi})}\,. 
\end{equation}
Interchanging the order of the $w$ and $\theta$ integrations, we can do the integral over $w$ first using the steepest descent method:
\begin{equation}
\int_0^{\infty} dw\, \cdot w^{d-1}\,\cdot  e^{q X(w)}\,\cdot e^{- q w f(\theta)} = \sqrt{\frac{\pi}{q}} \frac{\cosh^{d-1}{(f/2)}}{\sinh^{2q+d}{(f/2)}}\,,
\end{equation}
noting that $Re(f)>0$. This leaves us with one final integration:
\begin{equation}
{\cal G}_E^{\rm wkb}(\cos{\phi}) \sim \int_0^{\pi} d\theta\, e^{\frac{d-1}{2}f(\theta)} \frac{\cosh^{d-1}{(f(\theta)/2)}}{\sinh^{2q+d}{(f(\theta)/2)}} \sin^{d-2}{\theta}\,.
\end{equation}
We then write
\begin{equation}
\frac{1}{\sinh^{2q+d}{(f(\theta)/2)}} = 2^{q+d/2} (\cosh{f(\theta)}-1)^{-d/2} e^{-q\ln{(\cosh{f(\theta)}-1)}}\,,
\end{equation}
and switch to the complex variable $z=e^{i\theta}$ to yield
\begin{equation}
{\cal G}_E^{\rm wkb}(\cos{\phi}) \sim \int_C dz\,\cdot g(z)\cdot e^{q\,\chi(z)}\,.  
\end{equation}
Here
\begin{align}
g(z) &= \frac{1}{z}\left(z-\frac{1}{z}\right)^{d-2} e^{\frac{d-1}{2}f(z)} \frac{\cosh^{d-1}{\frac{1}{2}f(z)}}{(\cosh{f(z)}-1)^{d/2}}\,, \nonumber \\
\chi(z) & = -\ln{(\cosh{f(z)}-1)}\,, \nonumber \\
f(z) & = -\ln{(\cos{\phi}+\frac{i}{2}(z+1/z)\sin{\phi})}\,,
\end{align}
and $C$ is the upper half of the unit circle centred on $z=0$ traversed counterclockwise.   The two saddle points where $\chi'(z)=0$ are at $z=1$ and $z=-1$.  The values of $\chi$ and $\chi''$ at these points are
\begin{align}
\chi(1) &= -\ln{(\cos{\phi}-1)},\qquad  \chi''(1) =-e^{-i\phi}(1+\cos{\phi})\nonumber \\
\chi(-1) &= -\ln{(\cos{\phi}-1)},\qquad  \chi''(-1) =-e^{i\phi}(1+\cos{\phi})
\end{align}
and $g(z)$ expanded around these points is
\begin{align}
g(z) &= \frac{2^{\frac{d-4}{2}}}{i^d} e^{-i\frac{d-1}{2}\phi} \frac{\cos^{d-1}{\frac{1}{2}\phi}}{\sin^d{\frac{1}{2}\phi}} (z-1)^{d-2}+... \nonumber \\
g(z) &= -\frac{2^{\frac{d-4}{2}}}{i^d} e^{i\frac{d-1}{2}\phi} \frac{\cos^{d-1}{\frac{1}{2}\phi}}{\sin^d{\frac{1}{2}\phi}} (z+1)^{d-2}+... 
\end{align}
Finally,  the WKB correlator comes out to be
\begin{equation}
\boxed{{\cal G}_E^{\rm wkb}(\cos{\phi}) \sim \frac{1}{\sin^{2q+d}{\frac{1}{2}\phi}}}\,,
\end{equation}
in agreement with the exact CFT result. {\em Therefore the  WKB approximation is exact in the conformal case.}

\end{document}